\RequirePackage{ifpdf}
\documentclass[11pt]{article}

\usepackage{jheppub}
\usepackage{multicol}
\usepackage{graphicx}
\usepackage{amsmath}
\usepackage{url}
\usepackage{multirow}
\usepackage{rotating}
\usepackage{afterpage}
\usepackage{lmodern}
\usepackage[T1]{fontenc}

\usepackage[table]{xcolor}
\usepackage{caption}
\usepackage{subcaption}
\usepackage{aas_macros}
\usepackage{jdefs}
\usepackage{bm}
\usepackage{ragged2e}

\usepackage{xspace}

\newcommand{\adelaide}{ARC Centre for Dark Matter Particle Physics, Department of Physics, University of Adelaide, Adelaide, SA 5005, Australia}
\newcommand{\ucla}{Physics and Astronomy Department, University of California, Los Angeles, CA 90095, USA}
\newcommand{\nanjing}{Department of Physics and Institute of Theoretical Physics, Nanjing Normal University, Nanjing, Jiangsu 210023, China}
\newcommand{\bom}{Bureau of Meteorology, Melbourne, VIC 3001, Australia}
\newcommand{\monash}{School of Physics and Astronomy, Monash University, Melbourne, VIC 3800, Australia}
\newcommand{\uq}{School of Mathematics and Physics, The University of Queensland, St.\ Lucia, Brisbane, QLD 4072, Australia}
\newcommand{\cambridge}{Cavendish Laboratory, University of Cambridge, JJ Thomson Avenue, Cambridge, CB3 0HE, UK}
\newcommand{\kicc}{Kavli Institute for Cosmology, Madingley Road, Cambridge, CB3 0HA, UK}
\newcommand{\uam}{Higher Polytechnic School, Universidad Autonoma de Madrid, Francisco Tomas y Valiente 25, Madrid, Spain}
\newcommand{\imapp}{High Energy Physics, IMAPP, Radboud University Nijmegen, Heyendaalseweg 135, 6525 AJ, Nijmegen, Netherlands}
\newcommand{\nikhef}{Nikhef, Science Park 105, 1098 XG Amsterdam, Netherlands}
\newcommand{\uniiceland}{Science Institute, University of Iceland, Dunhaga 7, IS-107 Reykjavik, Iceland}
\newcommand{\nordita}{Nordita, KTH Royal Institute of Technology and Stockholm University, Roslagstullsbacken 23, SE-106 91 Stockholm, Sweden}
\newcommand{\ific}{Instituto de F\'isica Corpuscular, IFIC-UV/CSIC, Carrer del Catedrátic José Beltrán Martinez, 2, Valencia, Spain}

\newcommand{\eindhoven}{Eindhoven University of Technology, Groene Loper 5, 5612 AZ Eindhoven, Netherlands}
\newcommand{\ox}{Rudolf Peierls Centre for Theoretical Physics, 20 Parks Road, Oxford OX1 3PU, United Kingdom}
\newcommand{\imperial}{Department of Physics, Imperial College London, Blackett Laboratory, Prince Consort Road, London SW7 2AZ, UK}
\newcommand{\grappa}{Gravitation Astroparticle Physics Amsterdam (GRAPPA), Institute for Theoretical Physics Amsterdam and Delta Institute for Theoretical Physics, University of Amsterdam, Science Park 904, 1098 XH Amsterdam, The Netherlands}

\newcommand{\abc}{\textsf{Artificial Bee Colony}\xspace}
\newcommand{\ampgo}{\textsf{AMPGO}\xspace}
\newcommand{\ampgofull}{\textsf{Adaptive Memory Programming for Global Optimisation}\xspace}
\newcommand{\bo}{\textsf{Bayesian Optimisation}\xspace}
\newcommand{\cmaes}{\textsf{CMA-ES}\xspace}
\newcommand{\cmaesfull}{\textsf{Covariance Matrix Adaptation Evolution Strategy}\xspace}
\newcommand{\de}{\textsf{Differential Evolution}\xspace}
\newcommand{\diver}{\textsf{Diver}\xspace}
\newcommand{\gambit}{\textsf{GAMBIT}\xspace}
\newcommand{\GB}{\gambit}
\newcommand{\github}{\textsf{GitHub}\xspace}
\newcommand{\gpbo}{\textsf{Gaussian Process-based Bayesian Optimisation}\xspace}
\newcommand{\gpyopt}{\textsf{GPyOpt}\xspace}
\newcommand{\gwo}{\textsf{Grey Wolf Optimisation}\xspace}
\newcommand{\hds}{\textsf{HDS}\xspace}
\newcommand{\hdsfull}{\textsf{High-Dimensional Sampling}\xspace}
\newcommand{\ide}{\textsf{iDE}\xspace}
\newcommand{\jde}{\textsf{jDE}\xspace}
\newcommand{\lambdajde}{\textsf{$\lambda$jDE}\xspace}
\newcommand{\particlefilter}{\textsf{Gaussian Particle Filter}\xspace}
\newcommand{\pso}{\textsf{Particle Swarm Optimisation}\xspace}
\newcommand{\pygmo}{\textsf{PyGMO}\xspace}
\newcommand{\pygmoabc}{\textsf{PyGMO Artificial Bee Colony}\xspace}
\newcommand{\pygmode}{\textsf{PyGMO Differential Evolution}\xspace}
\newcommand{\pygmogwo}{\textsf{PyGMO Grey Wolf Optimisation}\xspace}
\newcommand{\pyscannerbit}{\textsf{pyScannerBit}\xspace}
\newcommand{\randomsampling}{random sampling\xspace}
\newcommand{\scannerbit}{\textsf{ScannerBit}\xspace}
\newcommand{\trustregionbo}{\textsf{Trust Region Bayesian Optimisation}\xspace}
\newcommand{\turbo}{\textsf{TuRBO}\xspace}
\newcommand{\turbofull}{\textsf{Trust Region Bayesian Optimisation}\xspace}

 \newlength{\wth}
\title{A comparison of optimisation algorithms for high-dimensional particle and astrophysics applications}
\abstract{Optimisation problems are ubiquitous in particle and astrophysics, and involve locating the optimum of a complicated function of many parameters that may be computationally expensive to evaluate. We describe a number of global optimisation algorithms that are not yet widely used in particle astrophysics, benchmark them against \randomsampling and existing techniques, and perform a detailed comparison of their performance on a range of test functions. These include four analytic test functions of varying dimensionality, and a realistic example derived from a recent global fit of weak-scale supersymmetry. Although the best algorithm to use depends on the function being investigated, we are able to present general conclusions about the relative merits of \randomsampling, \de, \pso, the \cmaesfull, \bo, \gwo, and the \pygmoabc, \particlefilter and \ampgofull algorithms.}

\collaboration{The DarkMachines High Dimensional Sampling Group}  
\author[1]{Csaba~Bal\'azs,}
\author[2]{Melissa van Beekveld,}
\author[3,4]{Sascha~Caron,}
\author[5]{Barry~M.~Dillon,}
\author[6]{Ben~Farmer,}
\author[7]{Andrew~Fowlie,}
\author[8]{Eduardo~C.~Garrido-Merch{\'a}n,}
\author[9,10]{Will~Handley,}
\author[3,4]{Luc~Hendriks,}
\author[11,12]{Gu{\dh}laugur~J{\'o}hannesson,}
\author[13]{Adam Leinweber,}
\author[14]{Judita~Mamu{\v{z}}i{\'{c}},}
\author[15]{Gregory~D.~Martinez,}
\author[3,16]{Sydney Otten,}
\author[14]{Roberto Ruiz de Austri,}
\author[17,18]{Pat~Scott,}
\author[13]{Zachary Searle,}
\author[3,4]{Bob Stienen,}  
\author[19]{Joaquin Vanschoren,}
\author[13]{Martin~White}  

\affiliation[1]{\monash}
\affiliation[2]{\ox}
\affiliation[3]{\imapp}
\affiliation[4]{\nikhef}
\affiliation[5]{Institut f{\"u}r Theoretische Physik, Universit{\"a}t Heidelberg, Germany}
\affiliation[6]{\bom}
\affiliation[7]{\nanjing}
\affiliation[8]{\uam}
\affiliation[9]{\cambridge}
\affiliation[10]{\kicc}
\affiliation[11]{\uniiceland}
\affiliation[12]{\nordita}
\affiliation[13]{\adelaide}
\affiliation[14]{\ific}
\affiliation[15]{\ucla}
\affiliation[16]{\grappa}
\affiliation[17]{\uq}
\affiliation[18]{\imperial}
\affiliation[19]{\eindhoven}

\emailAdd{martin.white@adelaide.edu.au}
\emailAdd{bstienen@science.ru.nl}

\begin{document}

\maketitle


\section{Introduction}
Typical theories in various branches of science, such as particle physics, particle astrophysics and astrophysics, are formulated as parametric models.  To make predictions in these models, one needs to specify the values of a set of free numerical parameters.  Comparison with experiment, in turn, can constrain the values of these parameters and single out a theory that matches observation the best.  Such parameter estimation and model selection are fundamental components of the scientific method in the physical sciences that we hope will lead us to the best description of nature.

In the modern era, the exploding size of experimental data sets, coupled with complicated theories, has introduced a substantial computational complexity in parameter extraction and model selection. It is frequently necessary to perform computationally expensive simulations of experiments, and further problems result from the moderately large dimensionality of typical beyond-Standard Model physics models, whose parameter multiplicity often ranges up to $\mathcal{O}(100)$.
In parameter spaces of this size, sampling randomly or on a uniform grid is both inefficient and unlikely to lead to robust statistical conclusions \cite{AbdusSalam:2020rdj}.  

We have entered a period in fundamental physics where one experiment is unlikely to unambiguously determine the next theory of particle physics or cosmology. Consequently, we must combine clues from many different branches of observation. Fortunately, at the same time, exponentially increasing computing power has allowed scientists to become more ambitious both in the scope of observation and theoretical calculation. Thus, parameter extraction and model comparison can be performed at an extraordinary scale to find the theory that best describes the physical world, if one employs efficient and ingenious sampling algorithms. 

The likelihood function is a key quantity in statistical inference (see e.g., Ref.~\cite{Cousins:2020ntk} for a pedagogical introduction), as it tells us the probability of the observed experimental data for a particular set of model parameters. 
If we are considering data from several experiments, the likelihood function may often be written as a product of likelihoods, one for each of the individual experiments, if the individual experiments are independent. 
%
%
In frequentist statistics, one can obtain consistent estimators for the values of the parameters of the model by finding the set of parameters that maximises the likelihood, and use the maximum likelihood itself to construct a test-statistic to perform statistical tests (see e.g., Ref.~\cite{Cowan:2010js} for an introduction to likelihood-based tests in particle physics). The difficulty is that the likelihood function is rarely known as a simple function of the original parameters and so we cannot find the maximum analytically. In fact, although in our setting we assume a tractable likelihood, evaluating the likelihood may still involve non-differentiable forward simulations of experiments (see e.g., Ref.~\cite{Balazs:2017moi}) and so even derivatives are unavailable. We are thus forced to use derivative-free numerical optimisation algorithms (see Ref.~\cite{RiosReview} for a review) to explore the likelihood function.

Furthermore, likelihood functions of interest often contain multiple modes, i.e., several distinct local maxima. In this setting, exploring the likelihood function and locating the global maximum may be extremely challenging, as we risk getting stuck in a local maximum. For this reason, we focus on stochastic algorithms that, e.g., step out of a local maximum with a particular probability, and neglect local optimizers commonly used in physics~\cite{James:1975dr}.
The simplest such approach is repeated random sampling from the entire parameter space (followed by picking either the highest or lowest value found.). This, however, is known to be deficient for two reasons. First, as the dimension of the parameter space increases, the number of samples that need to be drawn increases exponentially if the same point density is to be maintained. In practice this would thus come with an exponentially higher demand for computational power, which is worsened by the fact that the function evaluations are typically already costly themselves. Second, it is highly inefficient in most physical examples, as the high likelihood regions of the parameter space usually occupy a very small region of the total multidimensional volume. 
The past decades have thus seen the development of a series of novel sampling and optimisation procedures, particularly metaheuristic ones, many of which have been utilised in particle astrophysics applications~\cite{Feroz:2011bj,Baltz04,Allanach06,SFitter,Ruiz06,Strege15,Fittinocoverage,Catalan:2015cna,MasterCodeMSSM10,2007NewAR..51..316T,2007JHEP...07..075R,Roszkowski09a,Martinez09,Roszkowski09b,Roszkowski10,Scott09c,Akrami09,BertoneLHCDD,Akrami:2010dn,SBCoverage,Nightmare,BertoneLHCID,Akrami11coverage,IC22Methods,SuperbayesXENON100,SuperBayesGC,Buchmueller08,Buchmueller09,MasterCodemSUGRA,MasterCode11,MastercodeXENON100,MastercodeHiggs,Buchmueller:2014yva,Bagnaschi:2016afc,Bagnaschi:2016xfg,Allanach:2007qk,Abdussalam09a,Abdussalam09b,Allanach11b,Allanach11a,Farmer13,arXiv:1212.4821,Fowlie13,Henrot14,Kim:2013uxa,arXiv:1503.08219,arXiv:1604.02102,Han:2016gvr, Bechtle:2014yna,arXiv:1405.4289,arXiv:1402.5419, MastercodeCMSSM, arXiv:1312.5233,arXiv:1310.3045,arXiv:1309.6958, arXiv:1307.3383, arXiv:1304.5526,arXiv:1212.2886, Strege13,Gladyshev:2012xq, Kowalska:2012gs,Mastercode12b,arXiv:1207.1839, arXiv:1207.4846, Roszkowski12,SuperbayesHiggs, Fittino12,Mastercode12, arXiv:1111.6098,Fittino, Trotta08,Fittino06,arXiv:1608.02489,arXiv:1507.07008, Mastercode15, arXiv:1506.02499,arXiv:1504.03260,Mastercode17,Cheung:2012xb,Arhrib:2013ela,Sming14,Chowdhury15,Liem16,LikeDM,Banerjee:2016hsk,Matsumoto:2016hbs,Cuoco:2016jqt,Cacchio:2016qyh,BertoneUED,Chiang:2018cgb,hepfit,Matsumoto:2018acr,1306.2144,0809.3437,0704.3704,Dunkley05,GreAT,Hastings,vanBeekveld:2019tqp}. 

The purpose of this paper is to survey a wide range of optimisation techniques that, to the best of our knowledge, have not received mainstream use in particle astrophysics applications.  
The different techniques are explored by different authors of this paper in the form of the following challenge: use your optimisation technique of choice to find the optima of a set of reference functions, including both analytic examples and a 12-dimensional parameter space representing a supersymmetric model called the phenomenological MSSM7 \cite{Athron:2017yua} (a popular theory of beyond-Standard Model Particle Physics). Apart from the common set of reference functions and the use of a common test framework, there was no common tuning of the free parameters of the techniques. Although this introduces a human factor in the experiments, we believe this is representative of a real-life application of any one of the explored methods. The work was completed within the DarkMachines community\footnote{\url{http://www.darkmachines.org}}, which aims to develop new approaches for dark matter research thorough closer collaboration with machine learning and data science experts. It is worth noting that the algorithms we compare of course have many uses beyond maximising likelihood functions, such as minimisation of fine-tuning or the  optimisation of the hyperparameters of an algorithm. Our ultimate aim is to provide a self-contained overview of optimisation methods that can be used as a reference by researchers working in the physical sciences. For the purposes of this publication, we developed a testing framework in Python with interfaces to codes representing each of the optimisation techniques we use below.

This paper is structured as follows.  In Section~\ref{sec:algorithms}, we define our optimisation problem in more detail, and provide a description of each of the techniques used.  We describe the test functions that we use for our comparative studies in Section~\ref{sec:funcdef}, and detail the results in Section~\ref{sec:results}.  Finally, we present conclusions in Section~\ref{sec:conclusions}, and describe our Python framework for implementing the scanning techniques in Appendix~\ref{app:python}. The best found solutions for each investigated function and algorithm, together with the corresponding algorithm's hyperparameters, can be found in Appendix~\ref{app:tables}.

\section{Optimisation}\label{sec:algorithms}

The problem that we address in this review is the following. Given a deterministic function $f:\mathbb{R}^n\rightarrow \mathbb{R}$, defined over a domain of interest given by lower and upper bounds on the function parameters, what is the optimum of the function? In particle astrophysics applications, the function can represent the likelihood of observed data given a particular physical theory, and the optimum then gives the maximum likelihood estimate of the parameters, which is an important quantity in frequentist inference. It is in fact more usual to minimise the negative log-likelihood function, rather than maximise the likelihood.

Optimisation techniques can be divided into categories, based on whether they require a knowledge of the derivatives of the function or not (either analytical or numerical). Since derivative information is not always available in the physical sciences, we focus on techniques that only require evaluations of the likelihood function itself. 
Arguably the most challenging optimisation problems in particle astrophysics applications arise in global fits of beyond-Standard Model physics models. Popular techniques for performing frequentist inference on dark matter models have included \emph{Markov Chain Monte Carlo} techniques (see e.g., Ref.~\cite{Hogg:2017akh}) and \emph{nested sampling}~\cite{Skilling:2006gxv} which, although designed for Bayesian computation, can be repurposed for frequentist studies~\cite{Feroz:2011bj,Baltz04,Allanach06,SFitter, Ruiz06,Strege15,Fittinocoverage,Catalan:2015cna,MasterCodeMSSM10,2007NewAR..51..316T,2007JHEP...07..075R,Roszkowski09a,Martinez09,Roszkowski09b,Roszkowski10,Scott09c,BertoneLHCDD,SBCoverage,Nightmare,BertoneLHCID,IC22Methods,SuperbayesXENON100,SuperBayesGC, Buchmueller08,Buchmueller09,MasterCodemSUGRA,MasterCode11,MastercodeXENON100,MastercodeHiggs,Buchmueller:2014yva,Bagnaschi:2016afc,Bagnaschi:2016xfg,Allanach:2007qk,Abdussalam09a,Abdussalam09b,Allanach11b,Allanach11a,Farmer13,arXiv:1212.4821,Fowlie13,Henrot14,Kim:2013uxa,arXiv:1503.08219,arXiv:1604.02102,Han:2016gvr, Bechtle:2014yna, arXiv:1405.4289, arXiv:1402.5419, MastercodeCMSSM, arXiv:1312.5233, arXiv:1310.3045, arXiv:1309.6958, arXiv:1307.3383, arXiv:1304.5526, arXiv:1212.2886, Strege13, Gladyshev:2012xq, Kowalska:2012gs, Mastercode12b, arXiv:1207.1839, arXiv:1207.4846, Roszkowski12, SuperbayesHiggs, Fittino12, Mastercode12, arXiv:1111.6098, Fittino, Trotta08, Fittino06,arXiv:1608.02489, arXiv:1507.07008, Mastercode15, arXiv:1506.02499, arXiv:1504.03260, Mastercode17,Cheung:2012xb,Arhrib:2013ela,Sming14,Chowdhury15,Liem16,LikeDM,Banerjee:2016hsk,Matsumoto:2016hbs,Cuoco:2016jqt,Cacchio:2016qyh,BertoneUED,Chiang:2018cgb,hepfit,Matsumoto:2018acr}. In recent years, \emph{genetic algorithms} and, to an even greater extent, \de, have proven capable of adequately exploring very complex likelihood functions in multiple theories of dark matter~\cite{Akrami09,Kvellestad:2019vxm,Hoof:2018ieb,Athron:2018hpc,Athron:2017yua,Athron:2017qdc,Athron:2018vxy}. In selecting techniques for this study, we have focused on global optimisers that are expected to provide comparable performance to \de. We include the \de implementation previously studied in~\cite{ScannerBit} in order to benchmark the performance of our newly-explored techniques. The full list of algorithms that we explore is as follows.

\subsection{Optimisation algorithms}
\subsubsection{\de}

\de [DE; \citenum{StornPrice95, Price05wholebook, DasSuganthan11, Price13}] is a population-based heuristic optimisation strategy belonging to the class of \textit{evolutionary algorithms}.  DE does not rely on derivatives of the function being optimised, and is often the algorithm of choice for highly multimodal or otherwise poorly-behaved objective functions.

DE consists of evolving a population of $NP$ individuals or `target vectors' $\{\mathbf{X}_i^g\},$ of specific points in the parameter space, for a number of generations. Here $i$ refers to the $i$th individual, and $g$ corresponds to the generation of the population.  The initial generation is generally selected randomly within the parameter intervals to be sampled.

One generation is evolved to the next via three main steps: mutation, crossover and selection.  The simplest variant of the algorithm is known as \textsf{rand/1/bin}; the first two parts of the name refer to the mutation strategy (random population member, single difference vector), and the third to the crossover strategy (binomial).

Mutation proceeds by identifying an individual $\mathbf{X}_i$ to be evolved, and constructing one or more donor vectors $\mathbf{V}_i$ with which the individual will later be crossed over. In the \textsf{rand/1} mutation step, three unique random members of the current generation $\mathbf{X}_{r1}$, $\mathbf{X}_{r2}$ and $\mathbf{X}_{r3}$ are chosen (with none equal to the target vector), and a single donor vector $\mathbf{V}_i$ is constructed as
\begin{equation}
\mathbf{V}_i = \mathbf{X}_{r1} + F(\mathbf{X}_{r2} - \mathbf{X}_{r3}),
\label{eq:mutation}
\end{equation}
with the scale factor $F$ being a parameter of the algorithm. A more general mutation strategy known as \textsf{rand-to-best/1} also allows some admixture of the current best-fit individual $\mathbf{X}_\mathrm{best}$ in a single donor vector,
\begin{equation}
\mathbf{V}_i = \lambda\mathbf{X}_\mathrm{best} + (1-\lambda)\mathbf{X}_{r1} + F(\mathbf{X}_{r2} - \mathbf{X}_{r3}),
\label{eq:randtobest}
\end{equation}
according to the value of another free parameter of the algorithm, $\lambda$.

Crossover then proceeds by constructing a trial vector $\mathbf{U}_i$, by selecting each component (parameter value) from either the target vector, or from one of the donor vectors.  In simple binomial crossover (the \textsf{bin} of \textsf{rand/1/bin}), this is controlled by an additional algorithm parameter $Cr$.  For each component of the trial vector $\mathbf{U}_i$, a random number is chosen uniformly between 0 and 1; if the number is greater than $Cr$, the corresponding component of the trial vector is taken from the target vector; otherwise, it is taken from the donor vector.  At the end of this process, a single component of $\mathbf{U}_i$ is chosen at random, and replaced by the corresponding component of $\mathbf{V}_i$ (to make sure that $\mathbf{U}_i\ne \mathbf{X}_i$).

Selection simply faces the target vector $\mathbf{X}_i$ off against the trial vector $\mathbf{U}_i$, with the vector returning the best value of the objective function retained for the next generation.  In this way, each member of a generation is pitted against exactly one trial vector in each generation step.

A widely-used variant of simple \textsf{rand/1/bin} \de is so-called \jde \cite{Brest06}, where the parameters $F$ and $Cr$ are optimised on-the-fly by the \de algorithm itself, as if they were regular parameters of the objective function. An even more aggressive variant known as \lambdajde \cite{ScannerBit} is the self-adaptive equivalent of \textsf{rand-to-best/1/bin}, where $F,$ $Cr$ and $\lambda$ are all dynamically optimised.  

In this paper, we run two different software implementations of \de. The first is the open-source implementation of the \lambdajde algorithm contained in the \diver package\footnote{\url{https://diver.hepforge.org}}. We use this via the \pyscannerbit interface to the \scannerbit package of the \GB code for beyond-Standard Model global statistical fits~\cite{ScannerBit,Athron:2017ard,Kvellestad:2019vxm}. We also run the \jde~\cite{jde} and \ide~\cite{ide} algorithms implemented in the \pygmo package~\cite{pygmo}. In doing so, we have varied the number of generations and the parameter adaptation scheme which is used to optimize the weight coefficient and the crossover probability. 

\subsubsection{\pso}

\pso [PSO; \citenum{488968, Bonyadi}] is another population-based evolutionary algorithm that does not make use of derivatives.  Here, each member of the population of parameter samples (`the swarm') is also given a velocity.  In each generation step, the positions of other particles in the swarm are used to update each particle's velocity.  The position of each particle is updated by allowing it to move along its velocity vector for a fixed amount of time.

The standard velocity update for particle $i$ in generation $g$ is
\begin{equation}
\mathbf{v}_i^{g+1} = \omega\mathbf{v}_i^g + \phi_1 r_1 (\mathbf{x}_{i,\mathrm{pb}}-\mathbf{x}_i^g) + \phi_2 r_2 (\mathbf{x}_{\mathrm{gb}}-\mathbf{x}_i^g),
\label{eq:velocity_update}
\end{equation}
such that $\mathbf{x}_i^{g+1} = \mathbf{x}_i^{g} + \mathbf{v}_i^{g}$.  Here $r_1$ and $r_2$ are uniform random numbers between 0 and 1, $\mathbf{x}_{i,\mathrm{pb}}$ is the $i$th particle's personal best-fit position so far (i.e.\ in any generation), $\mathbf{x}_{\mathrm{gb}}$ is the global best-fit position so far (i.e.\ by any particle, in any generation), and $\omega$, $\phi_1$ and $\phi_2$ are free parameters of the algorithm.

In this paper, we will make use of a self-adaptive variant inspired by \jde~\cite{jde} that we will refer to as \textsf{j-Swarm}, where $\omega$ and/or $\phi_1$ and $\phi_2$ can be dynamically optimised in the course of a run by treating them as parameters of the objective.  The implementation that we use is bundled in \scannerbit, and will be released within \GB and described in a forthcoming \GB publication.

\subsubsection{\cmaes}

The \textsf{Covariance Matrix Adaptation} (\textsf{CMA}) \textsf{Evolution Strategy} (\textsf{ES}) is another evolutionary optimisation algorithm based on the idea of natural selection \cite{Hansen:2006CMAES}.  From an initial point in the $n$--dimensional parameter space, $\mathbf{x}^{(0)}$, a set of $\lambda$ new points (called a population) are sampled from a multivariate normal distribution
\begin{equation}
    \mathbf{x}^{(g+1)}_k \sim \mathcal{N}\left(\mathbf{x}^{(g)},\left(\sigma^{(g)}\right)^2 \mathbf{C}^{(g)}\right)
    \label{eq:CMA_ES:update}
\end{equation}
with covariance matrix $\left(\sigma^{(g)}\right)^2 \mathbf{C}^{(g)}$, where $k=1,\ldots,\lambda$ and $g$ counts the number of generations.  The optimisation function is then evaluated at all $\mathbf{x}^{(g+1)}_k$. The obtained values are used to sort the points and the best $\mu$ points (the parents) used to calculate
\begin{equation}
    \mathbf{x}^{(g+1)} = \sum_{j=1}^\mu w_j x^{(g)}_j
    \label{eq:CMA_ES:weightsum}
\end{equation}
where $w_j>0$ are weights with $\sum_{j=1}^\mu w_j=1$ and $j$ is the sorted index running from best to worst point.

For optimum performance of the algorithm, the step size $\sigma^{(g)}$ and the matrix $\mathbf{C}^{(g)}$ should be updated to maximize the probability that the new generation is closer to the minimum of the objective function.  The optimal update has been found to be given with
\begin{equation}
\begin{split}
    \mathbf{p}_c^{(g+1)} ={}& (1-c_c)\mathbf{p}_c^{(g)} + \sqrt{c_c(2-c_c)\mu_{\text{eff}}} \, \frac{\mathbf{x}^{(g+1)}-\mathbf{x}^{(g)}}{\sigma^{(g)}},\\
    \mathbf{C}^{(g+1)} 
    ={}& (1-c_{\text{cov}}) \mathbf{C}^{(g)} + \frac{c_{\text{cov}}}{\mu_{\text{cov}}} \mathbf{p}_c^{(g+1)}\mathbf{p}_c^{(g+1)^T} \\
    &+ c_{\text{cov}} \left(1 - \frac{1}{\mu_{\text{cov}}}\right) \sum_{j=1}^{\mu} w_j \left(\frac{\mathbf{x}_j^{(g+1)}- \mathbf{x}^{(g)}}{\sigma^{(g)}}\right)\left(\frac{\mathbf{x}_j^{(g+1)}- \mathbf{x}^{(g)}}{\sigma^{(g)}}\right)^T.
    \end{split}
    \label{eq:CMA_ES:cm_update}
\end{equation}
Here, $\mathbf{p}_c^{(g)}$ is a cumulative path, storing information about the direction taken during previous steps, $c_c < 1$ is the learning rate for the cumulative path, $c_{\text{cov}} < 1$ is the learning rate for the covariance matrix and $\mu_{\text{cov}} \ge 1$ controls the ratio between cumulation and rank--$\mu$ updates.  The parameter $\mu_{\text{eff}}^{-1} = \sum_{j=1}^\mu w_j^2$ represents the effective selection mass.  The cumulation update adapts the matrix to the large scale gradient of the optimisation function, while the rank--$\mu$ update adapts to the local gradient.

The optimal update for the step size is based on the absolute length of the cumulative path. If consequent steps are taken in the same direction, the length is expected to be large, while for steps in random direction, the length is shorter.  In the former case, fewer and longer steps can be taken, while in the latter the step size should be decreased.  The update is therefore
\begin{equation}
    \begin{split}
    \mathbf{p}_\sigma^{(g+1)} &= (1-c_\sigma) \mathbf{p}^{(g)} + \sqrt{c_\sigma(2-c_\sigma)\mu_{\text{eff}}} \, \mathbf{C}^{(g)^{-1/2}} \left(\frac{\mathbf{x}^{(g+1)}-\mathbf{x}^{(g)}}{\sigma^{(g)}}\right)\\
    \sigma^{(g+1)} &= \exp{\left(\frac{c_\sigma}{d_\sigma}\left[\frac{\lvert \mathbf{p}_\sigma^{(g+1)}\rvert}{\lvert\mathcal{N}(0,\mathbf{I})\rvert}\right]\right)}
    \end{split}
    \label{eq:CMA_ES:sigma_update}
\end{equation}
with $\mathbf{C}^{(g)^{-1/2}} = \mathbf{B}^{(g)}\mathbf{D}^{(g)^{-1}}\mathbf{B}^{(g)^T}$ where $\mathbf{C}^{(g)} = \mathbf{B}^{(g)}\mathbf{D}^{(g)^{2}}\mathbf{B}^{(g)^T}$ is the eigenvalue decomposition of $\mathbf{C}^{(g)}$.  The adaptation speed is controlled by the learning rate $c_\sigma$ and the damping parameter~$d_\sigma$.

The \cmaes is an invariant and stationary optimisation algorithm, meaning that its tuning parameters are insensitive to the objective function and depend almost exclusively on the dimensionality of the parameter space.  Only a brief overview of the algorithm has been given here, for details and explanations see Ref.~\cite{Hansen:2006CMAES}.  The implementation used in this work is from the \textsf{pycma} package.\footnote{\url{https://github.com/CMA-ES/pycma}}

\subsubsection{\bo}
\bo [BO; \citenum{SnoekPBO, BullCGO, RasmGPML}] is a set of techniques that attempts to find the optimum $\mathbf{x}^*$ of an objective function $f(\mathbf{x})$ with the minimum number of function evaluations, which is particularly useful when the function is computationally expensive to evaluate. It works by explicitly approximating the objective function $f(\mathbf{x})$ with a probabilistic regression model, called a surrogate model, that can predict the outcome of yet unseen samples to make a more informed decision of which samples to evaluate next. The initial surrogate model is trained on a set of random samples of the objective function, or a set of samples selected by any other sampling technique. The surrogate model needs to be probabilistic, and popular choices are Gaussian processes or probabilistic ensembles. Every further sample of the objective function $f(\mathbf{x})$ counts as a training point $\mathbf{x}$, continuously updating the surrogate model to a new posterior distribution that, after a given number of samples $\mathcal{D}\{(\mathbf{x}_i, y_i)\}$, gives our best belief of what the objective function $f(\mathbf{x})$ looks like. It can also provide some uncertainty at each point in the parameter space by using Gaussian likelihoods. An acquisition function $\alpha(\mathbf{x})$ is used to choose where to sample next, taking the latest posterior of the surrogate model as an input. The acquisition function $\alpha(\mathbf{x})$ is easy to evaluate and can be sampled with techniques such as Thompson sampling. In this way, cheap samples of the surrogate model are used to guide the sampling, rather than expensive samples of the objective function $f(\mathbf{x})$ itself. To avoid getting stuck in local minima, the acquisition function trades off exploration and exploitation, exploring regions of high expected outcome and those where the model shows high uncertainty, and hence requires more training data than is currently available. The result of this procedure is that fewer samples are needed to find the optimum $\mathbf{x}^*$ of the objective function $f(\mathbf{x})$, but more computation is required for predicting each next sample to try.  More formally, the method consists of the following steps.

\paragraph{Step 1: Define a surrogate model}
Assuming that we use a Gaussian process (GP), we also need to choose a prior distribution and a covariance function, or kernel, that defines the shape of the regression curves. The surrogate model can be written as $f(\mathbf{x}) \sim GP(\mu(\mathbf{x}),k(\mathbf{x},\mathbf{x}'))$. We can assume a normal prior with $\mu(\mathbf{x}) = 0$ without loss of generality. A popular choice for the kernel $k$ is the radial basis function, also known as the square exponential kernel, in which the length scale $\lambda$ and signal variance $\sigma^2$ control how flexible or flat the surrogate model can be:

\begin{equation}
k(\mathbf{x},\mathbf{x}’) = \sigma^2 \cdot \exp{\left(-\frac{||\mathbf{x}-\mathbf{x}'||^2}{2\lambda^2}\right)},\quad \lambda > 0.
\end{equation}

\paragraph{Step 2: Choose an acquisition function}
A number of acquisition functions are commonly used, each defining a specific way to trade off exploration and exploitation. Given a surrogate model trained on $n$ samples that can return, for every input $\mathbf{x}$, the predicted mean $\mu(\mathbf{x})$ and standard deviation $\sigma(\mathbf{x})$, as well as the current best value $f^*$ and a tuneable parameter $\psi$ which balances exploration and exploitation, we can describe the following popular choices:

\begin{enumerate}
    \item Maximum probability of improvement (MPI):
    \begin{equation} \label{eq:bo_mpi}
        a_\text{MPI}(\mathbf{x}) = \Phi(\gamma(\mathbf{x})),\quad \textrm{ where } \gamma(\mathbf{x}) = \frac{\mu(\mathbf{x}) - f^* - \psi}{\sigma(\mathbf{x})}
    \end{equation}

    \item Expected improvement (EI):
    \begin{equation} \label{eq:bo_ei}
        a_\text{EI}(\mathbf{x}) = (\mu(\mathbf{x}) - f^*)\Phi(\gamma(\mathbf{x})) + \sigma(\mathbf{x}) \phi(\gamma(\mathbf{x}))
    \end{equation}

    \item Upper confidence bound (UCB):
    \begin{equation} \label{eq:bo_ucb}
        a_\text{UCB}(\mathbf{x}) = \mu(\mathbf{x}) - \psi\sigma(\mathbf{x})
    \end{equation}
\end{enumerate}
In these equations $\Phi$ and $\phi$ are the CDF and PDF of the standard normal distribution, respectively.

The acquisition function $\alpha(\mathbf{x})$ is sampled using, for instance, Thompson sampling, and the sample with the highest acquisition score will be evaluated next on the actual optimisation function. After evaluation, this sample becomes a new training point, the surrogate model is updated, and the next sample is selected. This process is iterated until an acceptable result is reached, a certain budget is exhausted, or when all acquisition scores fall below a predefined threshold. The final recommendation of the optimisation process is the best observed result $\mathbf{x}^*$ or the optimisation of the mean of the updated posterior distribution on all observations $\mathcal{D}\{(\mathbf{x}_i, y_i)\}$.

\subsubsection{\turbofull}
Standard \bo suffers from scalability issues in high-dimensional problems. This is mainly due to the implicit homogeneity assumption of most surrogate models, and an overemphasis on exploration in the used acquisition functions. The acquisition function $\alpha(\mathbf{x})$, also becomes difficult to optimize in high-dimensional problems as it has the same number of dimensions as the number of dimensions of the input space.

For instance, the commonly used Gaussian process surrogates in \bo assume a constant length scale $\lambda$ and signal variance $\sigma$ in the search space. This is often not the reality of high dimensional functions, which tend to be flat in most of the space between local or global optima. Moreover, in high-dimensional spaces, samples are few and far between, meaning that the surrogate will exhibit high uncertainty and cause the acquisition function $\alpha(\mathbf{x})$ to focus predominantly on exploration instead of exploitation. This will harm the performance of \bo.

To overcome those issues, the \turbofull (\turbo) algorithm \cite{eriksson2019scalable} fits a set of local models and determines how to allocate samples from those such that the global optimum $\mathbf{x}^*$ is found most efficiently. It performs a collection of simultaneous local optimisation executions using independent Gaussian processes. 

Through this procedure, each Gaussian process enjoys the typical benefits of Bayesian modeling and these local surrogates allow for heterogeneous modeling of the objective function $f(\mathbf{x})$ without suffering
from over-exploration. In order to optimize all surrogates, \turbo leverages an implicit multi-armed bandit strategy at each iteration to allocate samples between these local areas and thus decide which local optimisation runs to continue.

Each Gaussian process resides in a Trust Region, a hyperrectangle centered at the best solution found so far, $\mathbf{x}^*$, under a surrogate model with a base side length $L$ that is a function of the length scales $\lambda$ of the Gaussian process: $L_i = \lambda_i L / (\prod_{j=1}^{d} \lambda_j)^{1/d}$. The local optimisation runs use a batch acquisition function that is restricted to select points $\mathbf{X}$ that lie within the Trust Region hyperrectangle. 

The base side length $L$ evolves during the optimisation process. If $L$ contains all the input space $\mathcal{X}$ the algorithm becomes standard \bo. The trade-off between exploration, large $L$, and exploitation of good solutions, small $L$, becomes critical. A popular criterion is to double $L$ in size after $\tau$ results better than $\mathbf{x}^*$, and halve the size in the other case. 

\turbo maintains $m$ trust regions simultaneously, selecting \texttt{max\_eval} candidates drawn from the union of all trust regions and updating all local optimisation problems for which candidates were drawn. \turbo gets the $i^{\rm th}$ candidate from all the trust regions by drawing a sample of the posterior mean function from all the Gaussian Processes per trust region and getting the minimum from them all, which is a greedy Thompson Sampling approach: 

\begin{equation}
x_i = \min_l \min_{x \in {\rm trust\, region}} f_l \,,
\end{equation}

\noindent where $f_l$ is a sample from the GP corresponding to trust region $l$.

\subsubsection{\gwo}

\noindent The \gwo algorithm \cite{gworef} falls under the category of swarm intelligence algorithms, and takes inspiration from the hunting mechanism and leadership hierarchies in packs of grey wolves.  Like other swarm intelligence algorithms there are two main phases: exploration and exploitation. The algorithm takes inspiration from the behaviour of grey wolves here through their tracking and attacking of prey, in which the social hierarchy of the pack of grey wolves also plays an important role. In the grey wolf analogy the wolves are associated with the candidate solutions in the swarm and the prey is associated with the best solution to the optimisation problem. The search agents in the algorithm are assigned to one of four categories, $\alpha$, $\beta$, $\delta$, or $\omega$, which are meant to imitate the social hierarchy of the grey wolves. The fittest solution is assigned to $\alpha$, the second fittest to $\beta$, the third fittest to $\delta$, and the rest to $\omega$.  During the optimisation the search for the optimal solution is usually led by the $\alpha$ solution, however the $\beta$ and $\delta$ also have influence.

The algorithm is initiated by assigning the search agents to random solutions in the search space, and these candidate solutions are assigned to the categories based on their fitness. The positions of each search agent are then updated, with the updates containing stochastic components as well as influence from the positions of the $\alpha$, $\beta$, and $\delta$ search agents.  In order to model the search strategy of the search agents the following definitions are required:
\begin{align}
\vec{A}&=2\vec{a}\cdot \vec{r}_1 - \vec{a} \nonumber \\
\vec{C}&=2\vec{r}_2.
\end{align}
These vectors have a length equal to the dimension of the search space.  The vectors $\vec{A}$ and $\vec{C}$ will be used to update the positions of the search agents, where $\vec{r}_{1,2}$ and vectors of random numbers in $[0,1]$, and $\vec{a}$ has components which decrease linearly from $2$ to $0$ over the course of the iterations.  Vectors capturing the distance between each agent and the three fittest agents are defined as:
\begin{align}
D_{\alpha}&=|\vec{C}_1\cdot \vec{X}_{\alpha}-\vec{X}| \nonumber \\
D_{\beta}&=|\vec{C}_2\cdot \vec{X}_{\beta}-\vec{X}| \nonumber \\
D_{\delta}&=|\vec{C}_3\cdot \vec{X}_{\delta}-\vec{X}|.
\end{align}
The positions of each agent are then updated according to:
\begin{align}
\vec{X}(t+1)=\frac{\vec{X}_1+\vec{X}_2+\vec{X}_3}{3}
\end{align}
where $t$ labels the current iteration, and:
\begin{align}
\vec{X}_1&=\vec{X}_{\alpha}-\vec{A}_1\cdot D_{\alpha} \nonumber \\
\vec{X}_2&=\vec{X}_{\beta}-\vec{A}_2\cdot D_{\beta} \nonumber \\
\vec{X}_3&=\vec{X}_{\delta}-\vec{A}_3\cdot D_{\delta}.
\end{align}
The form of these updates are such that the $\alpha$, $\beta$, and $\delta$ agents estimate the position of the optima by encircling it, and the $\omega$ agents search randomly around this position.  The sequence of updates is continued until some condition on the fitness of the optimal solution is met, or for a fixed number of iterations.  The only hyper-parameters in this algorithm are the number of agents to use, which should be at least $4$.  It is possible to introduce other hyper-parameters by varying the possible range of the numbers in the vector $\vec{a}$ also, but this is not part of the minimal set-up described in \cite{gworef}.

\subsubsection{\pygmoabc}

The \abc algorithm is inspired by the intelligent behaviour of honey bees in their search for the right food sources \cite{abc, AKAY2012120}. The algorithm has an automated mechanism to balance exploration and exploitation. In the experiments performed for this paper we use the implementation of the \abc in \pygmo~\cite{pygmo}, for which the pseudocode can be found in Listing 2 of Ref.~\cite{MERNIK2015115}. Here we restrict ourselves to a more conceptual explanation of the algorithm.

The \abc algorithm keeps track of $SN$ active data points $x_s$, where $s$ runs from 1 to $SN$. The locations of these data points are uniformly initialised and the function value $f_s$ for each of these data points is evaluated. In the bee-analogy these data points can be seen as the locations of food sources and the function value can be interpreted as being related to the food gain from each of these sources. The goal of the algorithm is then to find the food source (i.e. data point) with the highest gain (i.e. best function value).

Finding this data point is done iteratively, where each iteration consists of three phases. In the first phase each active data point $x_{i}$ is used as reference to explore a new location $v_{i}$. The proposal location $v_{i}$ is calculated by moving, for each dimension $j$, in the direction of another, randomly selected active data point $x_{k}$:

\begin{equation} \label{eq:abc_combine}
    v_{i,j} = x_{i,j} + \varphi_{ij}(x_{i,j} - x_{k,j}),
\end{equation}

\noindent where $\varphi_{ij}$ is a uniform random number between 0 and 1. For the proposal $v_{i}$ the functional value $f_s'$ is calculated and if this value is better than $f_s$, $x_{i}$ and $f_s$ will be replaced by $v_{i}$ and $f_s'$ respectively. The old location and function value are kept if this test fails.

For each data point the number of failed update attempts is kept track of. This information can be used to give up on locations $x_{i}$ that fail to be updated for many subsequent iterations. This is done in the second phase, where these dead data points (as determined with respect to some user-configured threshold) are reinitialised uniformly in the parameter space.

These first two steps can be seen as a form of exploration: the parameter space is explored by sampling new data points and evaluating their function values. The third -- and last -- step of each iteration is a form of exploitation, in which Equation~\ref{eq:abc_combine} is used to perform updates on the active data points. Which active data points are updated is determined by assigning to each data point $x_{i}$ a fitness-score\footnote{The fitness-score presented here is the one for minimisation problems.}. This score is dependent on the functional value $f_i$ of each data point:

\begin{equation} \label{eq:abc_fitness}
    {\rm fitness}_i = \begin{cases}
        \left(1 + f_i\right)^{-1} & f_i \geq 0 \\
        1 + \left|f_i\right|      & f_i < 0 \\
    \end{cases}.
\end{equation}

\noindent This score is then used to calculate an update probability for each data point

\begin{equation} \label{eq:abc_probability}
    p_i = \frac{{\rm fitness}_i}{\sum_i {\rm fitness}_i}.
\end{equation}

\noindent The update attempts are distributed over the active data points using Bernoulli experiments with respect to these probabilities. Just like in the first phase data points are only updated if the proposal point $v_{i}$ improves on the function value of the original location $x_{i}$. Also in this phase the number of update attempts is kept track of.

The automated balancing of exploration on the one hand and exploitation on the other makes the \abc algorithm a potentially strong optimisation algorithm. Moreover, it has the nice property that as the number of iterations increases, data points will tend to move towards minima, because updates are only performed if they improve the function value for the data point under consideration. This causes the updates of Equation~\ref{eq:abc_combine} to provide increasingly more resolution on these minima.




\subsubsection{\particlefilter}

The \particlefilter was first explored in Ref.~\cite{1232326}. The scanning algorithm starts off by collecting an initial seed of randomly generated points. The number of points that are generated is a user definable quantity, as are the range and the sampling prior (uniform or logarithmic). By then using these randomly generated points as seeds, a multi-dimensional Gaussian distribution is used to draw new points around the location of those seed data points. The number of data points sampled from each Gaussian is proportional to the relative function value of the target function at the seed. The width of these Gaussians varies over the run time of the algorithm, starting at ($\mathcal{O}(1-2)$) at the start of the run, in order to cover a large part of the parameter space. In higher iterations, this width is shrunk by multiplying by a factor $<$ 1. The speed at which it reduces is a hyperparameter of the algorithm and will henceforth be called the width decay.

The implementation of the algorithm that we employ in this paper\footnote{\url{https://github.com/bstienen/particlefilter}} allows for the use of a uniform or logarithmic exploration, indicating if the standard deviations of the multi-dimensional Gaussian distributions should be multiplied by the coordinate $x$. This is configured through the boolean parameter \texttt{logarithmic}.

In each iteration $N_{\rm sample}$ data points are sampled from $N_{\rm Gaussians}$ Gaussians and the function values for these samples are calculated. These samples are combined with the best fraction of the (already evaluated) samples that formed the seed for the Gaussians. This fraction is named the survival rate.  From the resulting dataset, $N_{\rm Gaussians}$ data points (i.e. those with the lowest function value) are selected as the seed for the Gaussians of the next iteration.

The algorithm is highly flexible, and is aided by adding user knowledge about how the function behaves. This can help to determine an appropriate value to choose for the width of the Gaussians. The stopping criterion of the implementation that we employed was based entirely on the width of the Gaussians, which we controlled using a width scheduling scheme. We did not use any information on the function values found, allowing the algorithm to potentially continue to sample after already finding the global minimum.

\subsubsection{\ampgo}

\begin{figure}
    \centering
    \includegraphics[width=0.4\textwidth]{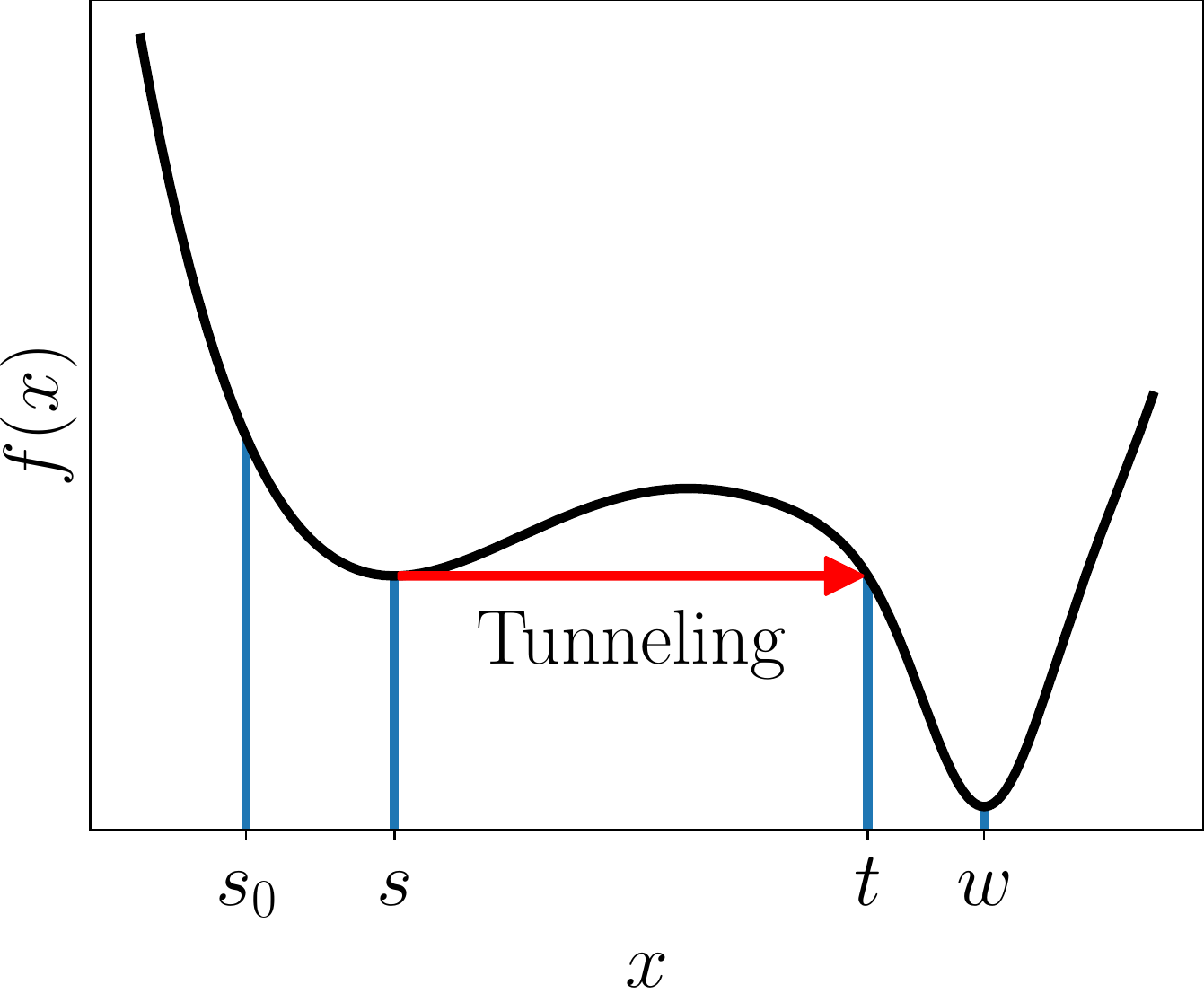}
    \caption{Visualisation of the tunneling approach for minimisation as used in AMPGO \cite{AMPGO}. From the initial sample $s_0$ the local optimum at $s$ is found. After it is found, Tabu Tunneling is used to find location $t$, from which the local gradient-based approach can be reused again to find the global minimum at $w$.}
    \label{fig:ampgo_tunneling}
\end{figure}

\ampgofull (\ampgo) is a solver that uses multiple steps to solve a global optimisation problem~\cite{AMPGO}. First, a set of initial points in parameter space is chosen randomly. Next, a local solver is used to find the local optima of those points. When these are found, a method called Tabu Tunneling \cite{AMPGO_TabuTunneling} is used to find a different local optimum in the parameter space from which the local optimizer is started again. Using this iterative approach, the global optimum can be found. In Figure \ref{fig:ampgo_tunneling} a visualisation of this approach is shown. The goal is to find the global optimum $w$ and the initial random point is $s_0$. As a first step, using the local optimizer, the point $s$ is found. Then, using the Tabu Tunneling method, one tries to find a different point in parameter space with a value as good as $s$. In this example, that point is $t$. From $t$ the local optimisation algorithm is invoked again to find the point $w$.

Any method can be used for the local solver; we employed \textsf{L-BFGS-B} \cite{Byrd94alimited-memory}. A full mathematical description of the algorithm can be found in Ref.~\cite{AMPGO}. 
The \ampgo website \cite{AMPGO_Site} contains benchmarks of 184 multidimensional test functions against multiple different algorithms.
A Python implementation of \ampgo is available on \github\footnote{\url{https://github.com/andyfaff/ampgo}}.

The algorithm relies heavily on tunneling towards a new region in parameter space with the same or lower function value. However, as the number of possible tunnelling directions from any point in parameter space is infinite, there are significant hurdles to overcome when trying to find a region with a lower or equal function value:

\begin{itemize}
    \item When the global optimum is very narrow, the probability of tunnelling into the basin of attraction of this minimum can be very low (e.g. Analytic Function 1 discussed in Section~\ref{sec:hidden});
    \item When the dimensionality increases, the volume to tunnel through in order to find a new minimum grows exponentially;
    \item When there are many local minima of similar depth, the number of tunneling steps needed can be very high. Each time a new local minimum is found that is deeper than the last, the probability of finding a new minimum with an even lower function value goes down (see e.g.\ Analytic Function 2).
\end{itemize}

Given these difficulties, this algorithm can struggle to improve on its local solution in high-dimensional problems.



\subsection{Characterisation of algorithms}

It is not immediately obvious how to compare optimisation algorithms that depend on wildly different strategies, with each having a number of different hyperparameters. The best choice of the hyperparameters in each case will depend on the function being explored, and finding those best values is itself an optimisation challenge of reasonable complexity. 

In this paper we attempt to provide the average particle astrophysicist with some knowledge of which algorithm might perform well for a particular type of function, without requiring them to engage in overly-aggressive hyperparameter optimisation. We therefore group the parameters of each of our techniques into four categories:

\begin{itemize}
    \item \textbf{Convergence parameters: }These are parameters that affect the stopping point of an algorithm, or the point at which convergence is presumed to have occurred. A more stringent tolerance condition should have the effect of improving the best fit found by the algorithm, but will require a greater number of likelihood evaluations to reach that point.
    \item \textbf{Resolution parameters: }These are parameters that affect the resolution with which the target function is explored. A higher resolution will increase the detail with which the likelihood function is mapped around the best-fit point, leading to a better mapping of the 1$\sigma$, 2$\sigma$ and 3$\sigma$ confidence regions, at the cost of a greater number of likelihood evaluations in total.  Higher resolution can also improve the final quality of the best-fit point, independently of convergence parameters. 
    \item \textbf{Hint parameters: }These are parameters that give the algorithm a clue as to how to obtain the best solution. For example, algorithms that are required to start at a certain point would strongly benefit from starting near the global minimum of the function. A wise choice of such a parameter (if this is possible) would reduce the number of likelihood evaluations required to give a good fit.
    \item \textbf{Reliability parameters: }These are parameters whose general effect is to improve the robustness of a technique.
\end{itemize}

In Table~\ref{tab:params}, we provide a grouping of the main parameters of each of our optimisation techniques, where it can be seen that not all techniques have a parameter in each group. Nevertheless, most of the techniques have a resolution and a convergence parameter, and the typical use-case for a particle astrophysicist would be to set these parameters to provide the best possible sampling within the available CPU budget, whilst using out-of-the-box choices for the hint and reliability parameters. Our results in Section~\ref{sec:results} will therefore be presented for different choices of the resolution and convergence parameters for each algorithm (where these are available). 

\begin{table}[]
    \centering
    \begin{tabular}{|p{5cm}p{7cm}r|}
        \hline
         Parameter & Explored values & Type \\
        \hline
        \multicolumn{3}{|l|}{\cellcolor{gray!25} \ampgo} \\
        Number of sampled points & 2000, 5000, 10000, 20000 & Resolution \\

        \multicolumn{3}{|l|}{\cellcolor{gray!25} \cmaes} \\
        Function tolerance & $10^{-11}$, $10^{-7}$, $10^{-4}$, $10^{-1}$ & Convergence \\
        Population size ($\lambda$) & $20$, $50$, $100$, $500$ & Resolution \\
        
        \multicolumn{3}{|l|}{\cellcolor{gray!25} \diver} \\
        Threshold for convergence & $10^{-4}$, $10^{-3}$, $10^{-2}$, $10^{-1}$ & Convergence \\
        Population size & 2000, 5000, 10000, 20000 & Resolution \\
        Parameter adaptation scheme & \lambdajde & - \\

        \multicolumn{3}{|l|}{\cellcolor{gray!25} \particlefilter} \\
        Width decay & 0.90, 0.95, 0.99& Convergence \\
        Logarithmic sampling & {\tt True}, {\tt False} & Hint \\
        Survival rate & 0.2, 0.5 & Reliability \\
        Initial gaussian width & 2 & Reliability \\
        
        \multicolumn{3}{|l|}{\cellcolor{gray!25} \gpyopt} \\
        Threshold for Convergence & $10^{-6}$, $10^{-5}$, $10^{-4}$, $10^{-3}$, $10^{-2}$, $10^{-1}$ & Convergence \\
        
        \multicolumn{3}{|l|}{\cellcolor{gray!25} \pso} \\
        Threshold for convergence & $10^{-4}$, $10^{-3}$, $10^{-2}$, $10^{-1}$ & Convergence \\
        Population size & 2000, 5000, 10000, 20000 & Resolution \\
        Adaptive $\phi$ & {\tt True} & Reliability \\
        Adaptive $\omega$ & {\tt True} & Reliability \\
        
        \multicolumn{3}{|l|}{\cellcolor{gray!25} \pygmoabc} \\
        Generations & 100, 250, 500, 750 & Resolution \\
        Maximum number of tries & 10, 50, 100 & Reliability \\
        
        \multicolumn{3}{|l|}{\cellcolor{gray!25} \pygmode} \\
        Generations & 100, 250, 500, 750 & Resolution \\
        Parameter adaptation scheme & \ide, \jde & - \\
        
        \multicolumn{3}{|l|}{\cellcolor{gray!25} \pygmogwo} \\
        Generations & 10, 50, 100, 1000 & Resolution \\
        
        \multicolumn{3}{|l|}{\cellcolor{gray!25} \randomsampling} \\
        Number of points & 10, 50, 100, 500, 1000, 5000, 10000, 50000, 100000, 500000, 1000000 & Resolution \\
        
        \multicolumn{3}{|l|}{\cellcolor{gray!25} \turbofull (\turbo)} \\
        Max \#evaluations / iteration & 64, 100 & Convergence \\
        \hline
    \end{tabular}
    \caption{A grouping of the various parameters of each of our optimisation techniques into the categories described in the main text. The explored values for these parameters can be found in the second column.}
    \label{tab:params}
\end{table}

\section{Definition of sampler comparison tests}
\label{sec:funcdef}
\subsection{Analytic test functions}
\label{sec:hidden}

In this paper two sets of tests are performed. In the first, we compare the performance of each of the above algorithms on a series of analytic test functions, each of which embodies a different pathology that is commonly asociated with difficult physics examples. In the second, a real physics example is used, in order to determine whether the insights gained on analytic test functions generalise to a more realistic setting.

We use four reference functions\footnote{Our choice of these functions is greatly influenced by~\cite{AMPGO_Site}.} to evaluate and compare the performance of every optimisation algorithm. The analytic forms of these functions were unknown to the operators of the optimisation algorithms until the results were generated, in order not to introduce biases towards particular global minima. 

\begin{figure}
    \centering
    \begin{subfigure}{0.49\textwidth}
        \centering
        \includegraphics[width=\textwidth]{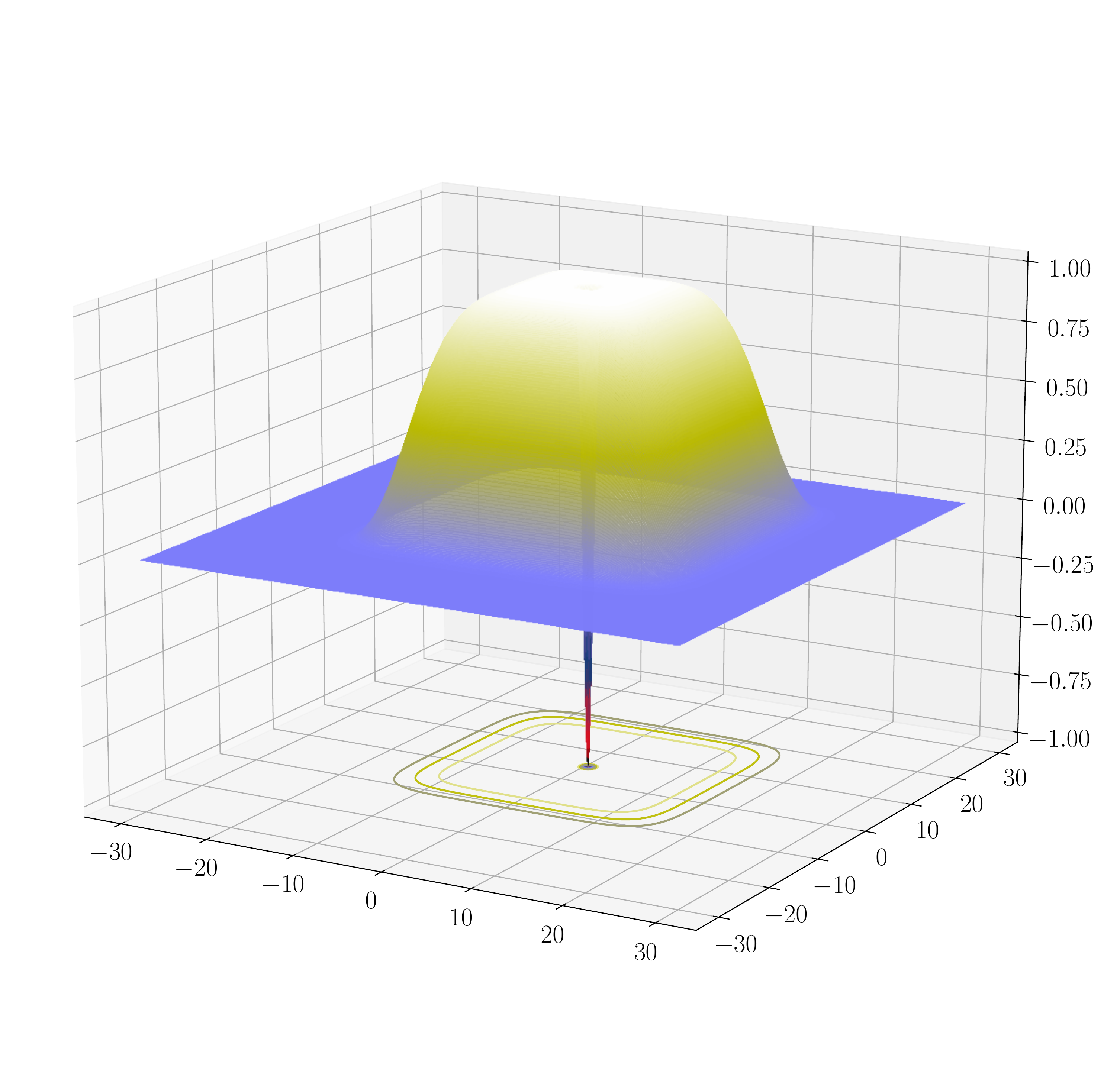}
        \caption{Analytic Function 1 (eq.~\ref{eq:hidden_1}) --- global minimum at $\mathbf{2}$.}
        \label{fig:hiddenfunction-XinSheYang03}
    \end{subfigure}
    \hfill
    \begin{subfigure}{0.49\textwidth}
        \centering
        \includegraphics[width=\textwidth]{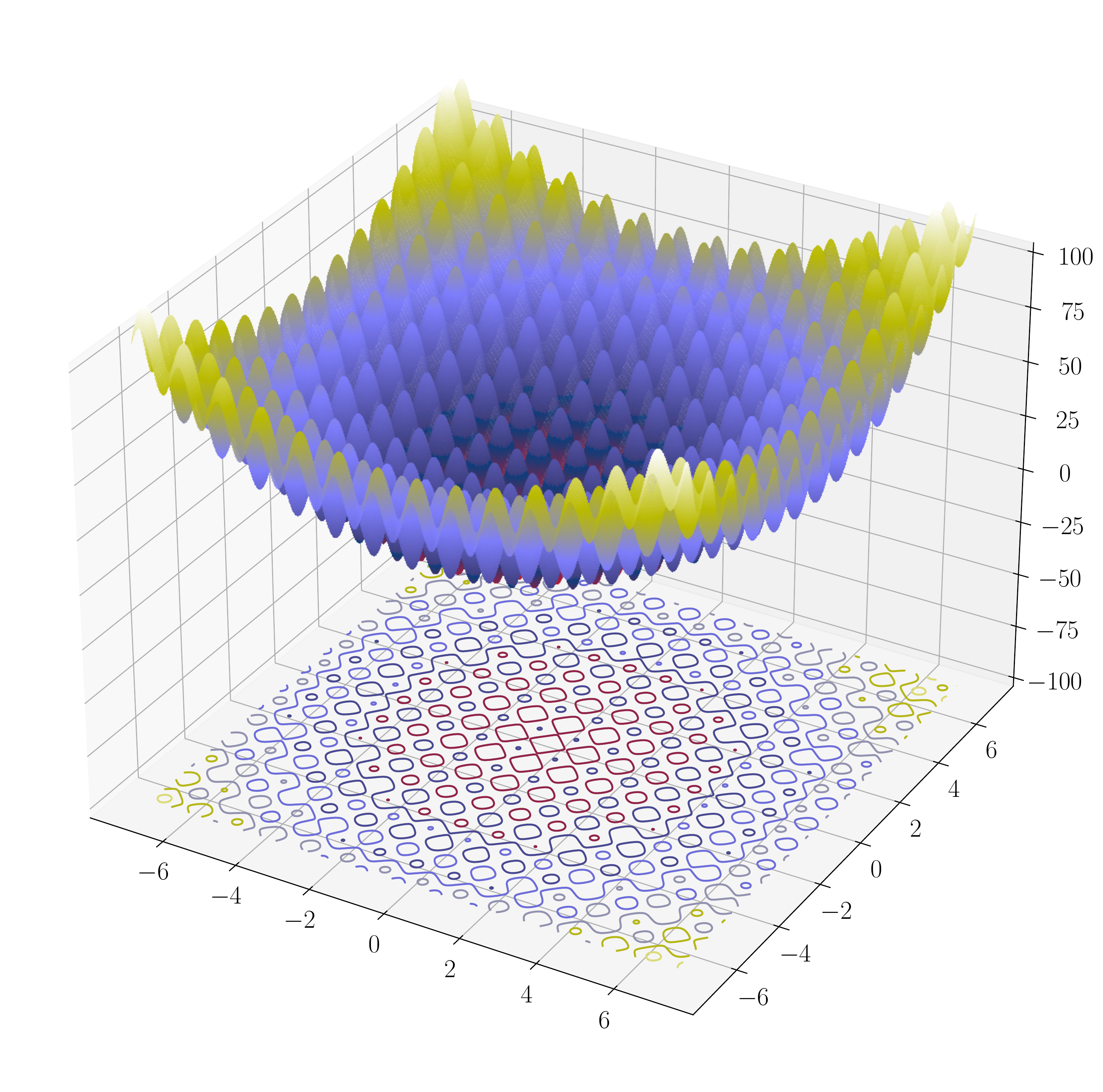}
        \caption{Analytic Function 2 (eq.~\ref{eq:hidden_2}) --- global minimum at $\mathbf{-0.23}$.}
        \label{fig:hiddenfunction-YaoLiu09}
    \end{subfigure}
    \hfill
    \begin{subfigure}{0.49\textwidth}
        \centering
        \includegraphics[width=\textwidth]{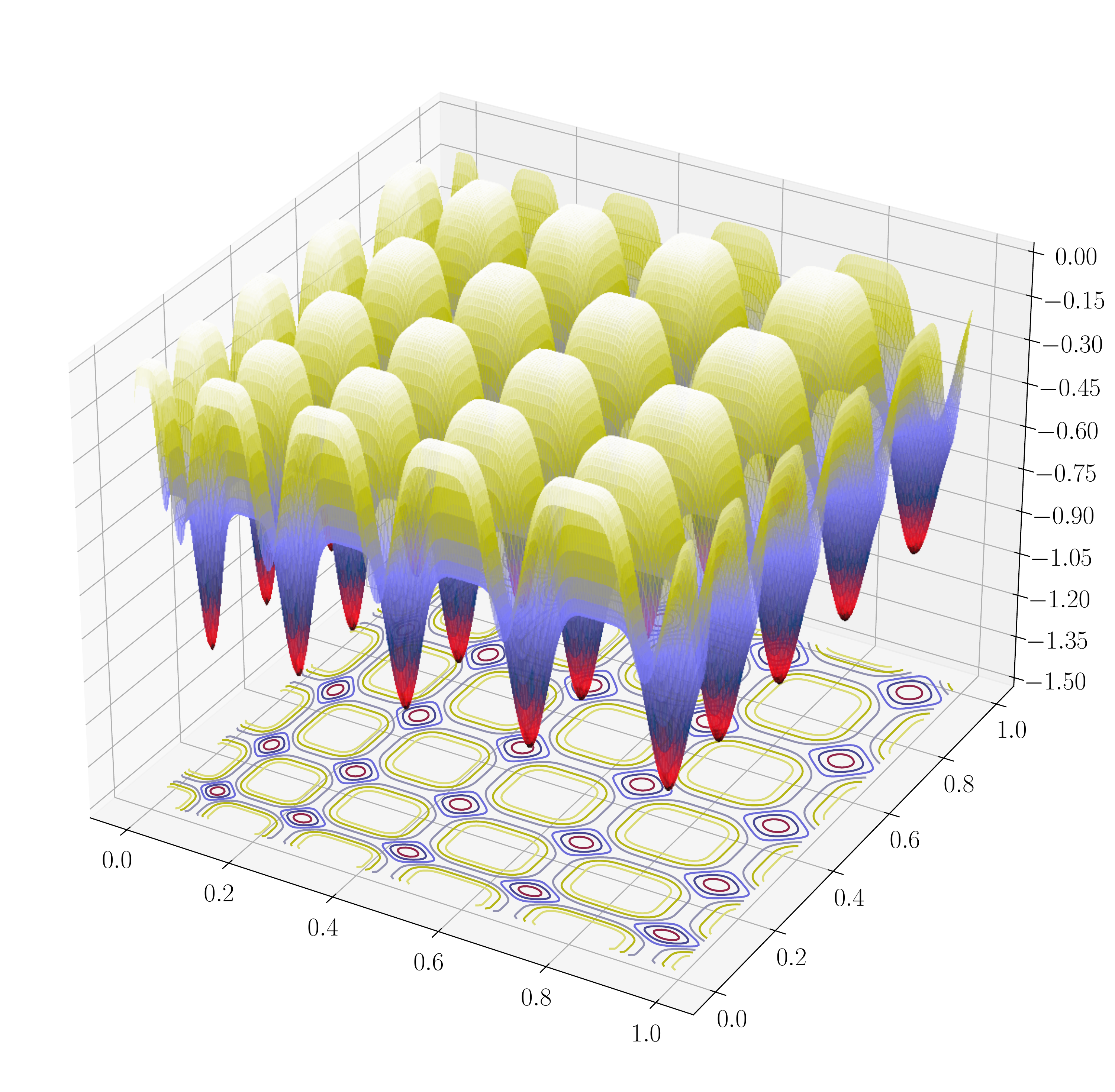}
        \caption{Analytic Function 3 (eq.~\ref{eq:hidden_3}) --- many degenerate minima.}
        \label{fig:hiddenfunction-Deb02}
    \end{subfigure}
    \hfill
    \begin{subfigure}{0.49\textwidth}
        \centering
        \includegraphics[width=\textwidth]{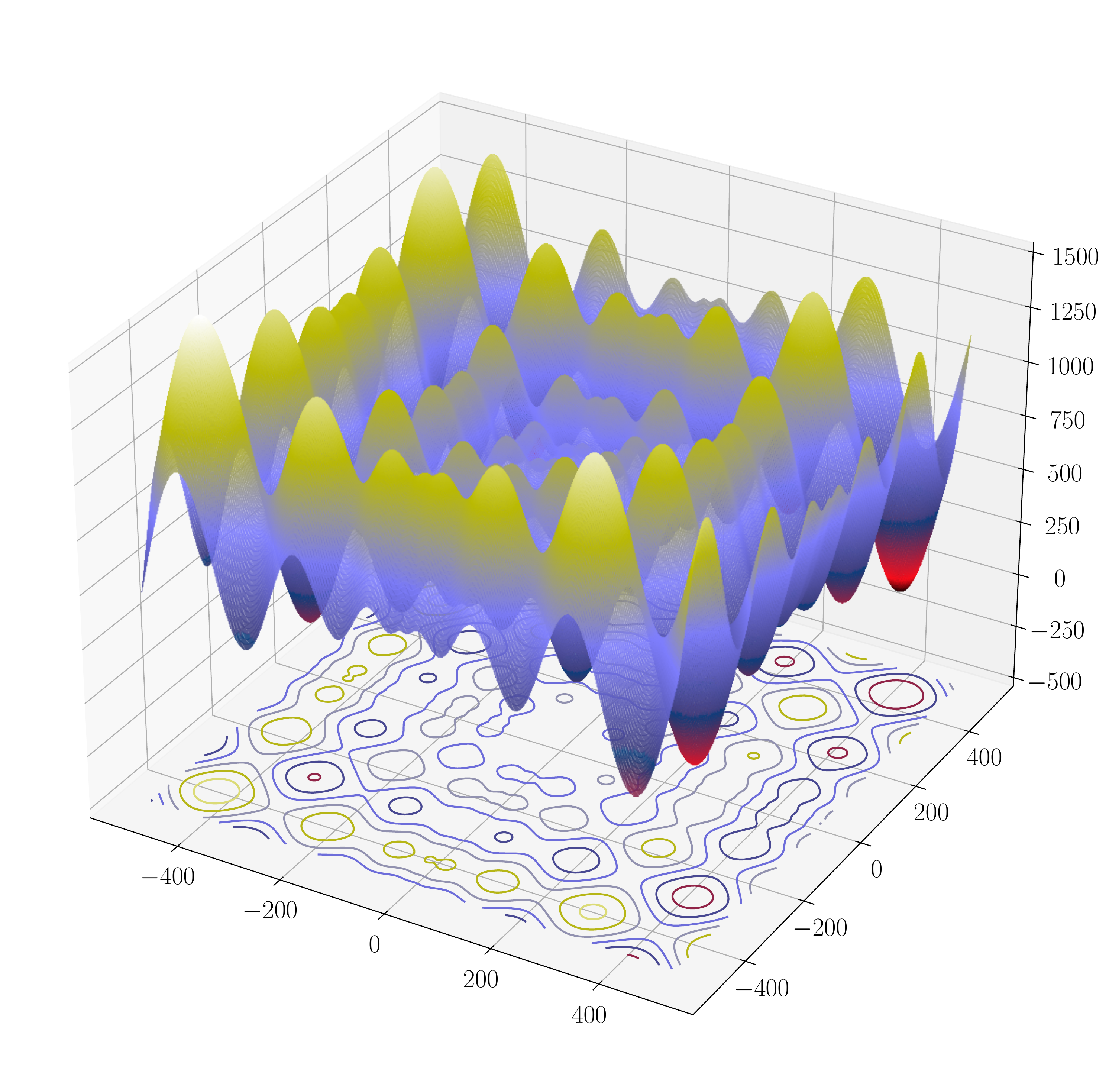}
        \caption{Analytic Function 4 (eq.~\ref{eq:hidden_4}) --- global minimum at about $\mathbf{421}$.}
        \label{fig:hiddenfunction-Schwefel26}
    \end{subfigure}
    \hfill
    \caption{Visualisation of the explored analytic functions from Section~\ref{sec:hidden} in 2-dimensional form.}
    \label{fig:hiddenfunctions}
\end{figure}

\subsubsection*{Analytic Function 1}
The equation for Analytic Function 1 reads as follows:

\begin{equation} \label{eq:hidden_1}
    f(\mathbf{x}) = e^{-\sum_{i=1}^n\left((x_i - 2)/15\right)^{6}} - 2e^{-\sum_{i=1}^n (x_i - 2)^2} \prod_{i=1}^n \cos^2 (x_i - 2),
\end{equation}

\noindent where $n$ is the number of dimensions. This function is visualised in Figure \ref{fig:hiddenfunction-XinSheYang03} for 2-dimensional input. The global minimum is at $\mathbf{2}$, where a function value of $-1$ is reached. This function is expected to be difficult to optimise numerically, as this minimum is surrounded by a region with relatively high function values. The global minimum is put at $\mathbf{2}$ rather than $\mathbf{0}$ to discourage algorithms that take zero as a starting point. The minimum of this function is searched for in the domain $[-30.0, 30.0]^n$.

\subsubsection*{Analytic Function 2}
The equation for Analytic Function 2 reads as follows:

\begin{equation} \label{eq:hidden_2}
    f(\mathbf{x}) = \sum_{i=1}^n\left[ (x_i + 0.23)^2-10\cos\left(2\pi (x_i + 0.23)\right) + 10 \right],
\end{equation}

\noindent where $n$ is the number of dimensions. This equation is visualised in Figure \ref{fig:hiddenfunction-YaoLiu09} for 2-dimensional input. The difficulty in this analytic function is that there are many local minima that the optimisation algorithm can get stuck in. The global minimum is put at $\mathbf{-0.23}$ rather than $\mathbf{0}$ to discourage algorithms that take zero as a starting point, with $f(\mathbf{-0.23})=0$. The minimum of this function is searched for in the domain $[-7.0, 7.0]^n$.

\subsubsection*{Analytic Function 3}
The equation for Analytic Function 3 reads as follows:

\begin{equation} \label{eq:hidden_3}
    f(\mathbf{x}) = -\frac{1}{n}\sum_{i=1}^n\sin^6\left[5 \pi \left(x_i^{3/4} -0.05 \right)\right],
\end{equation}

\noindent where $n$ is the number of dimensions. This equation is visualised in Figure \ref{fig:hiddenfunction-Deb02} for 2-dimensional input. This test function has many global minima and the optimisers should be able to find the global minimum, which is at $-1$. The minimum of this function is searched for in the domain $[0.0, 1.0]^n$.

\subsubsection*{Analytic Function 4}
The equation for Analytic Function 4 reads as follows:

\begin{equation} \label{eq:hidden_4}
    f(\mathbf{x}) = 418.9829n - \sum_{i=1}^n x_i \sin\left(\sqrt{|x_i|}\right),
\end{equation}

\noindent where $n$ is the number of dimensions. This equation is visualised in Figure \ref{fig:hiddenfunction-Schwefel26} for 2-dimensional input. The difficulty in this analytic function is that it has an irregular shape and the parameter range is quite large. The global optimum is at about $\mathbf{420.968746}$ with $f(\mathbf{420.968746}) \simeq 0$. The minimum of this function is searched for in the domain $[-500.0, 500.0]^n$.

\subsection{Particle astrophysics test problem} \label{ssec:mssm7}

In addition to the test functions described above, it is interesting to test our various optimisation algorithms on a realistic particle astrophysics problem. A leading use of optimisation techniques in particle astrophysics is in global fits of models beyond the Standard Model of particle physics. These add a number of additional parameters to the Standard Model, and one must find the regions of the extended parameter space that are most compatible with current experimental data. In frequentist statistics, this is typically performed by maximising a likelihood function $\mathcal{L}$, which is equivalent to minimising $-\log\mathcal{L}$. This problem therefore resembles the problem of minimising the analytic functions. 

For our example, we take a recent global fit of a supersymmetric theory performed by the \GB collaboration in Ref.~\cite{Athron_2017}, and obtain a fast interpolation of the likelihood function that was originally computationally expensive to obtain. The original fit explored a 7-parameter phenomenological version of the Minimal Supersymmetric Standard Model (the so-called ``MSSM7''), which is described by the soft masses $M_2$, $m^2_{\tilde{f}}$, $m^2_{H_u}$, $m^2_{H_d}$, the trilinear couplings for the third generation of quarks $A_{u_3}$, $A_{d_3}$ and tan $\beta$ (plus the input scale $Q=1$~TeV and the sign of $\mu$, which was chosen to be positive). The mass parameters above are all defined at the $Q$ common scale whereas tan $\beta$ is defined at $m_Z$. In addition to these supersymmetric model parameters, the original fit added a variety of nuisance parameters, comprising the strong coupling constant, the top quark mass, the local dark matter density, and the nuclear matrix elements for the strange, up and down quarks. Therefore, the global fit was performed on a 12-dimensional parameter space. 

To make it possible to compare the performance of the different algorithms considered in this work we have approximated the joint likelihood\footnote{See Tab. 3 of Ref.~\cite{Athron_2017} for details about which data were used to define it.} using a deep neural network as proposed e.g. in \cite{Brooijmans:2020yij,Buckley:2011kc}. The total number of samples collected from the global fit to train the network was about $2.3 \times 10^7$. The network consists out of four hidden layers of 20 fully-connected neurons each, activated through the SELU function \cite{klambauer2017selfnormalizing}. We have normalized the input and output data to a normal Gaussian distribution and split the data into 90\% for training and 10\% for testing. We have set the batch size for training to 1024 and optimized the performance by halving the learning rate of the Adam optimizer \cite{kingma2017adam} when the loss function, in this case the mean absolute error (MAE), stopped improving. Early stopping was applied to stop training after a couple of these iterations.

\begin{figure}[htbp]
    \centering
    \includegraphics[width=12cm]{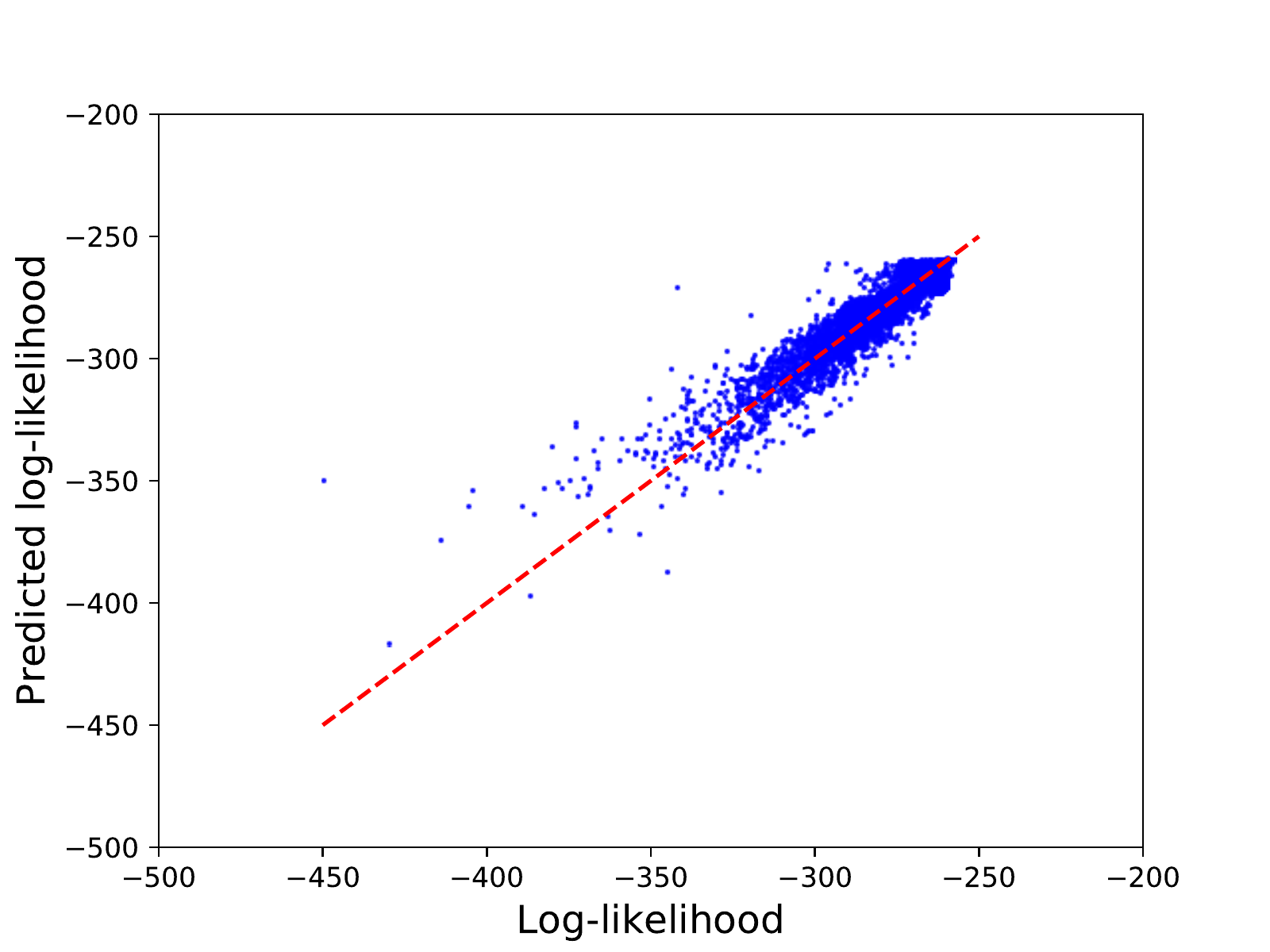}
     \caption{Validation plot of network, showing the true log likelihood on the $x$-axis and the predicted value on the $y$-axis.}
     \label{fig:mssm7}
\end{figure}

Figure~\ref{fig:mssm7} shows the validation plot of the trained network, where it can be seen that the neural net prediction of the likelihood is well-correlated with the true likelihood. It is important to note that perfect performance of the fast likelihood interpolation is not required, as for our purposes it is sufficient that it provides a suitable proxy for a difficult likelihood function that would typically be encountered in a particle astrophysics application.

\section{Results}
\label{sec:results}
In this section, we compare the results of running the optimisation methods described in Section~\ref{sec:algorithms} on the functions described in Section~\ref{sec:funcdef}. To make sure the algorithm was the only difference between all these optimisation experiments (apart from the algorithms' hyperparameters),  we performed all experiments using the same Python framework. A description of this framework can be found in Appendix~\ref{app:python}. A comprehensive overview of the best found result for each of the algorithms on each optimised function (for all explored dimensionalities) can be found in Appendix~\ref{app:tables}.

\subsection{Analytic Functions}
We first present results for the analytic test functions described in Section~\ref{sec:hidden}. These four functions were explored as 2-dimensional, 3-dimensional, 5-dimensional and 7-dimensional functions. It is expected that the ability of optimisation algorithms to find the true minimum decreases as the dimensionality increases. Each algorithm was run for several sets of its resolution and convergence hyperparameters, which are summarised in Table~\ref{tab:params}.

\begin{figure}
    \centering
    \includegraphics[width=0.9\textwidth]{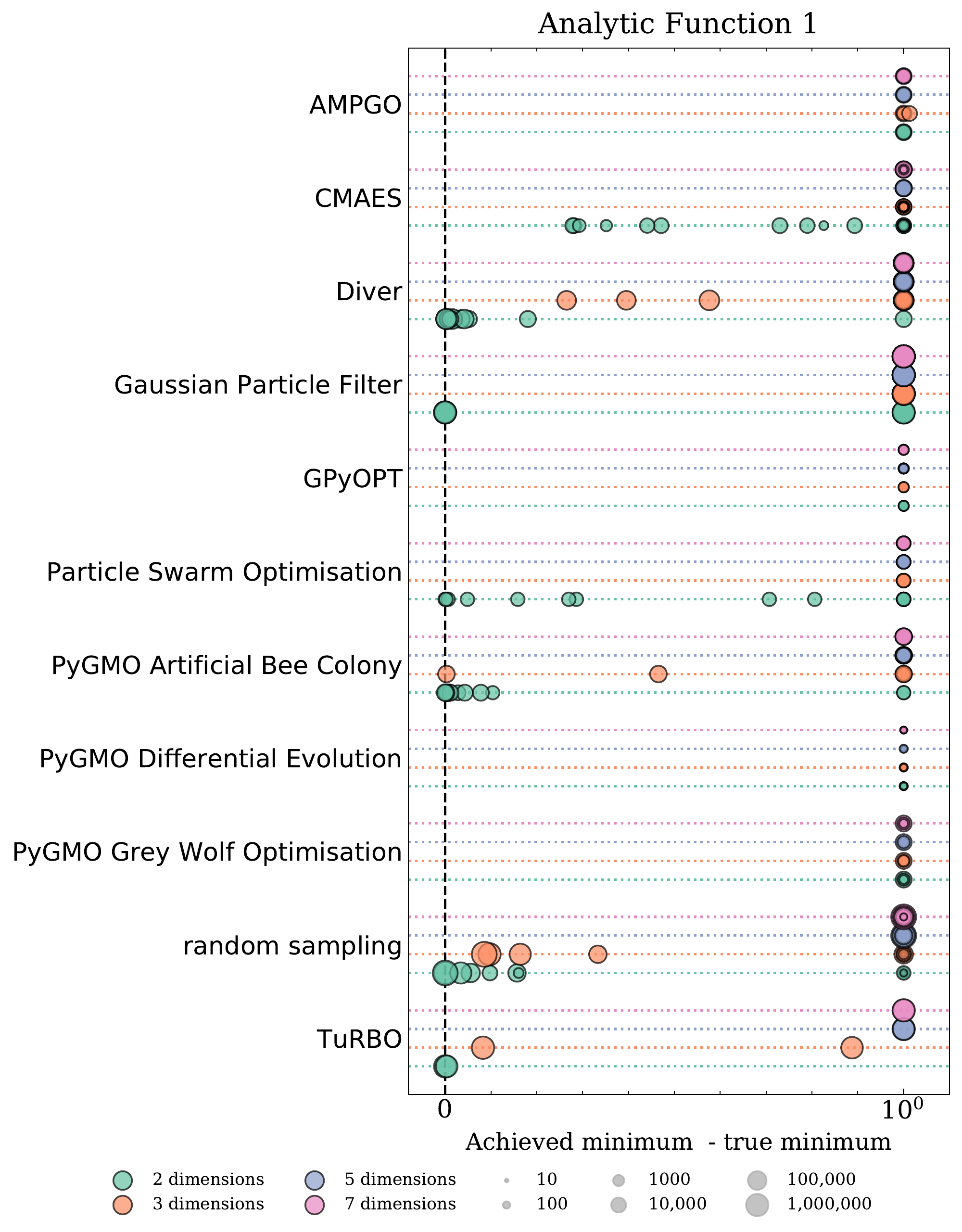}
    \caption{Results from different optimisation algorithms on the analytic function in Equation~\ref{eq:hidden_1}. The results are shown as semi-opaque circles, of which the area increases logarithmically with the number of function evaluations needed to obtain that specific result. The four horizontal lines for each algorithm belong to the four explored dimensionalities, from top to bottom 7-dimensional (pink), 5-dimensional (purple), 3-dimensional (orange) and 2-dimensional (green). The horizontal axis shows the difference between the (known) log-likelihood at the global minimum and that at the found minimum.}
    \label{fig:result_hidden_1}
\end{figure}

\begin{figure}
    \centering
    \includegraphics[width=0.9\textwidth]{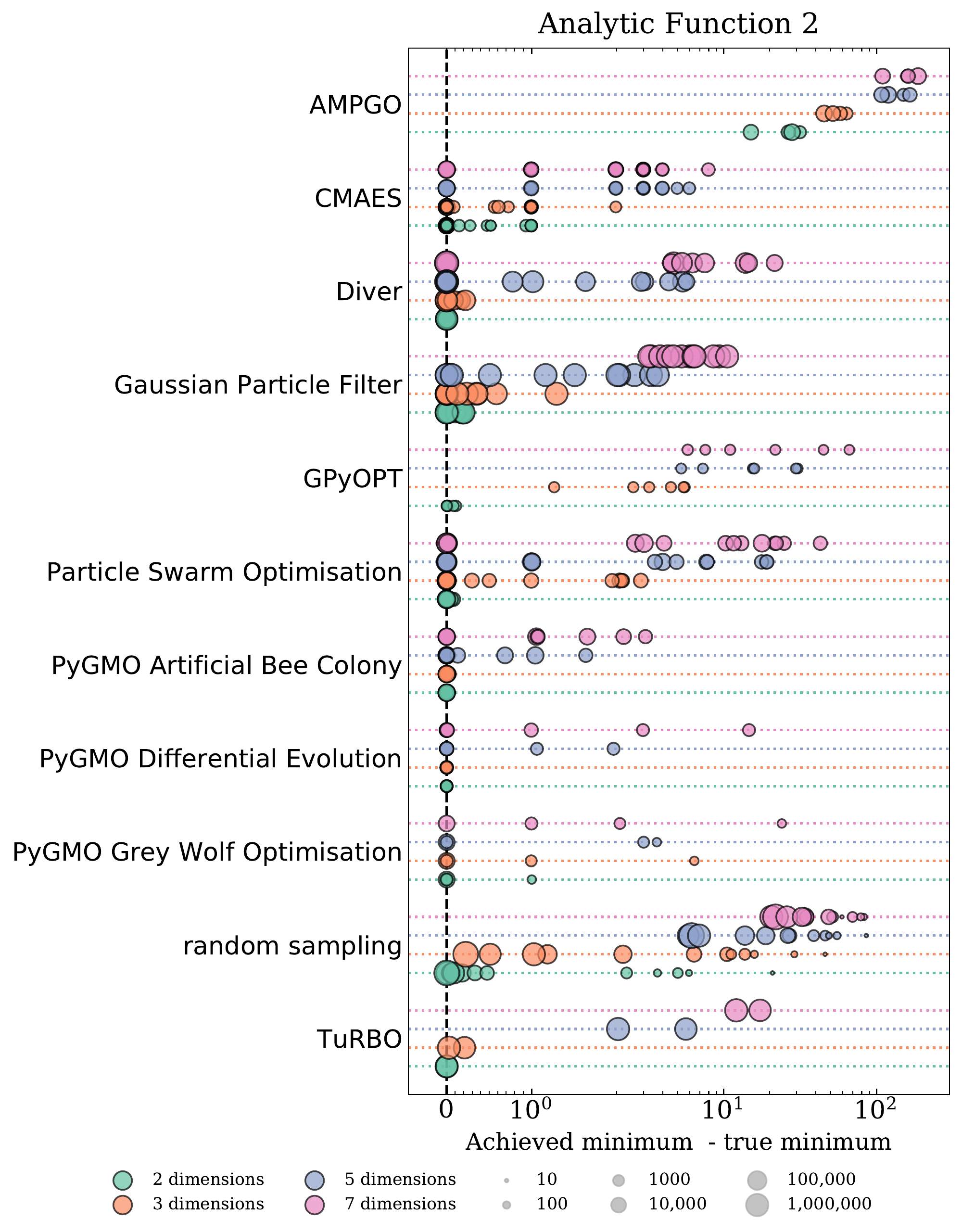}
    \caption{Results from different optimisation algorithms on the analytic function in Equation~\ref{eq:hidden_2}. The results are shown as semi-opaque circles, of which the area increases logarithmically with the number of function evaluations needed to obtain that specific result. The four horizontal lines for each algorithm belong to the four explored dimensionalities, from top to bottom 7-dimensional (pink), 5-dimensional (purple), 3-dimensional (orange) and 2-dimensional (green). The horizontal axis shows the difference between the (known) log-likelihood at the global minimum and that at the found minimum.}
    \label{fig:result_hidden_2}
\end{figure}

\begin{figure}
    \centering
    \includegraphics[width=0.9\textwidth]{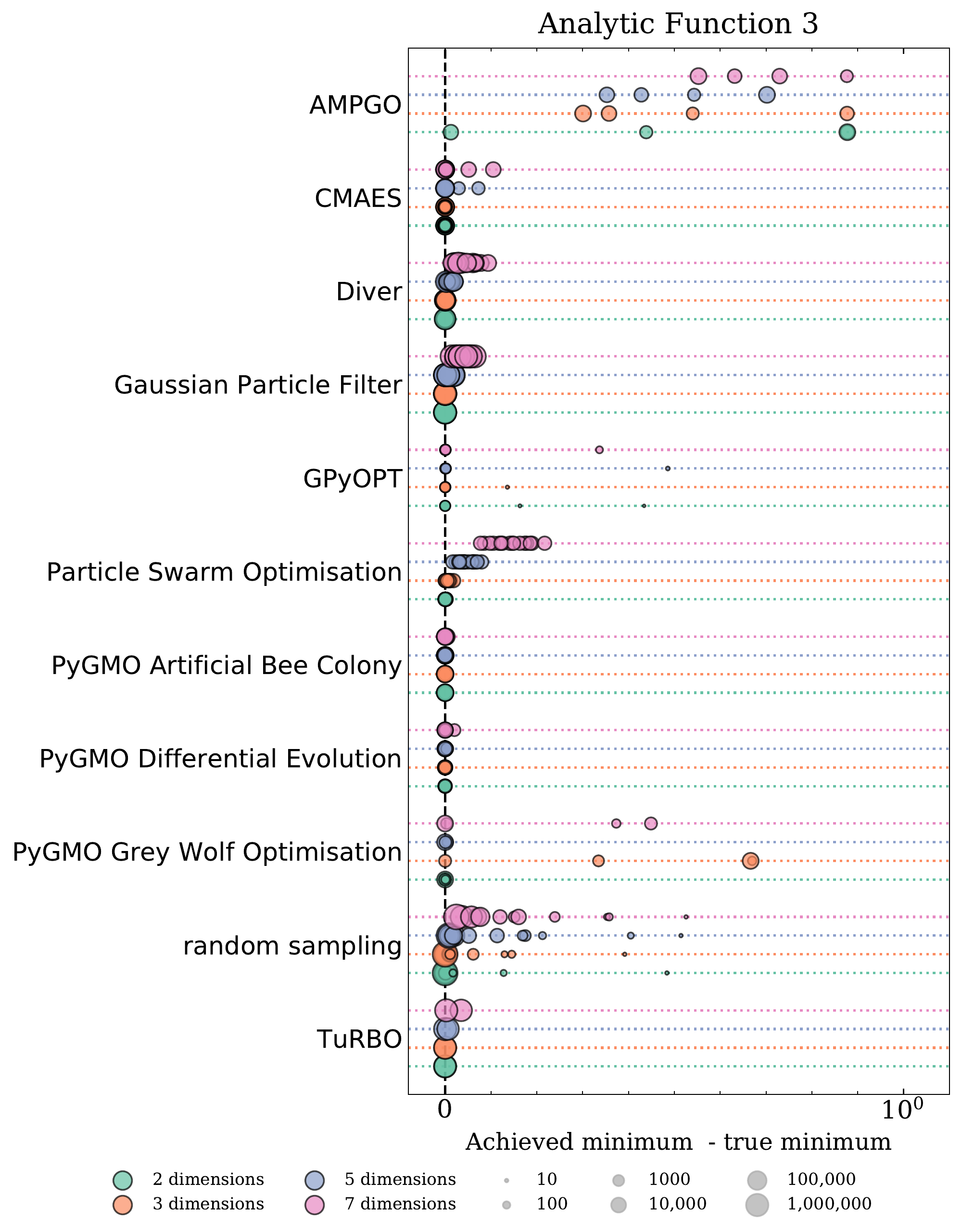}
    \caption{Results from different optimisation algorithms on the analytic function in Equation~\ref{eq:hidden_3}. The results are shown as semi-opaque circles, of which the area increases logarithmically with the number of function evaluations needed to obtain that specific result. The four horizontal lines for each algorithm belong to the four explored dimensionalities, from top to bottom 7-dimensional (pink), 5-dimensional (purple), 3-dimensional (orange) and 2-dimensional (green). The horizontal axis shows the difference between the (known) log-likelihood at the global minimum and that at the found minimum.}
    \label{fig:result_hidden_3}
\end{figure}

\begin{figure}
    \centering
    \includegraphics[width=0.9\textwidth]{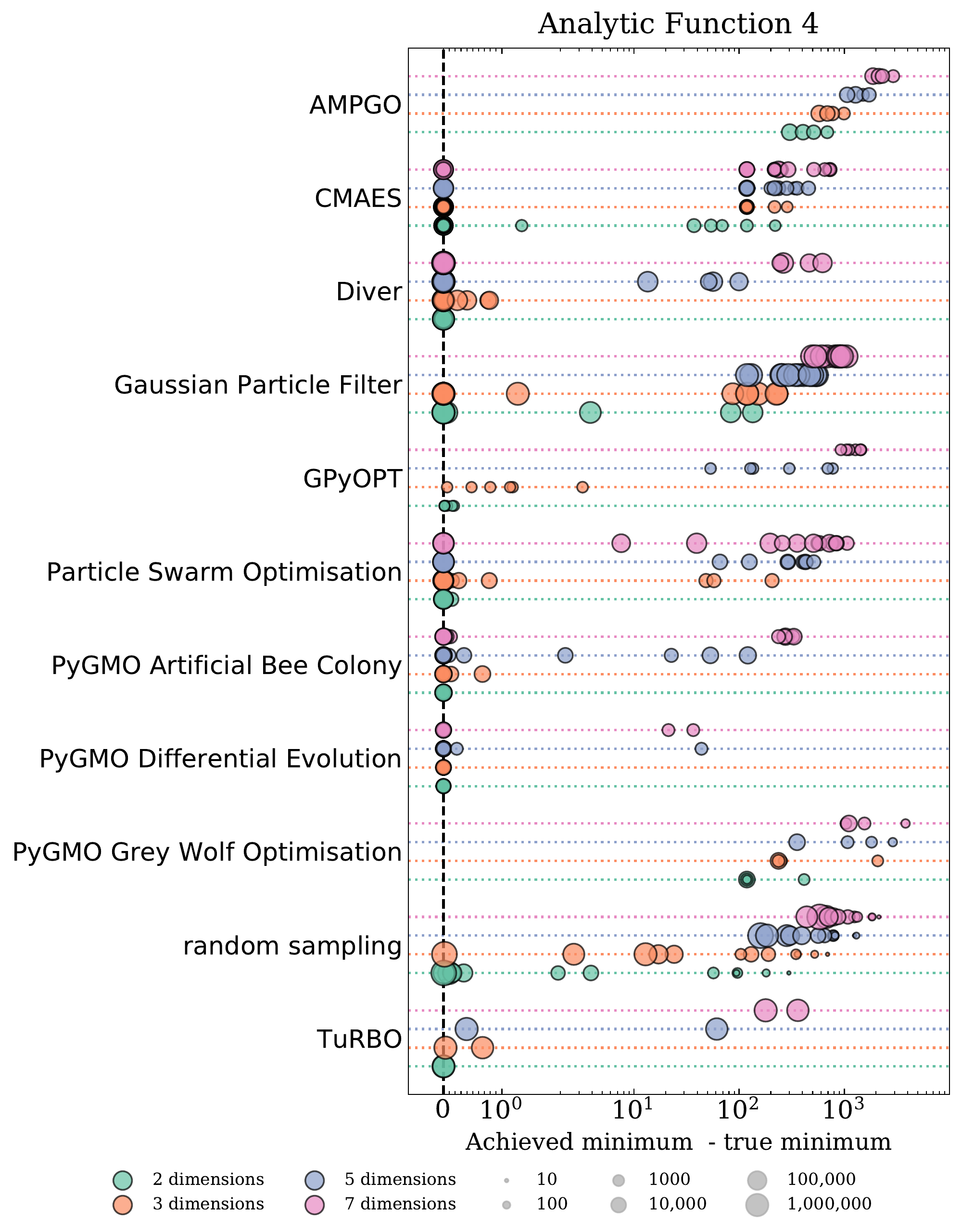}
    \caption{Results from different optimisation algorithms on the analytic function in Equation~\ref{eq:hidden_4}. The results are shown as semi-opaque circles, of which the area increases logarithmically with the number of function evaluations needed to obtain that specific result. The four horizontal lines for each algorithm belong to the four explored dimensionalities, from top to bottom 7-dimensional (pink), 5-dimensional (purple), 3-dimensional (orange) and 2-dimensional (green). The horizontal axis shows the difference between the (known) log-likelihood at the global minimum and that at the found minimum.}
    \label{fig:result_hidden_4}
\end{figure}

Figures~\ref{fig:result_hidden_1} to \ref{fig:result_hidden_4} show the accuracy with which each algorithm recovered the minimum of each function for each of the dimensionalities, with each circle representing a particular run of each algorithm with a specific choice of hyperparameters. The size of the circles is proportional to $\log_{10}$ of the total number of likelihood evaluations in that run, whilst the different lines for each algorithm correspond to different dimensionalities (going from 2D for the line on the bottom, to 7D for the top-most line). To illustrate the accuracy, we show the difference between the known global minimum and the best value found for each run. 


The results for Analytic Function 1 show that some algorithms never get anywhere close to the global minimum, regardless of the dimensionality of the problem. In $\ge 3$ dimensions, all algorithms fail to find the global minimum. Inspecting the function in 2D (see Figure~\ref{fig:hiddenfunction-XinSheYang03}) reveals why - there is a very spiked local minimum that is hard to locate, a problem that will get even harder as the dimensionality of the function increases. The best-performing algorithm, such as it is, is the \pygmoabc, which finds the correct minimum in 2D and 3D with a relatively small number of likelihood evaluations. The worst performing algorithms are \pygmogwo, \particlefilter, \ampgo, \gpyopt and the \pygmo implementation of \de.  In the latter case this may simply be due to the low number of total likelihood evaluations, suggesting that a more stringent set of hyperparameters might yield better performance. This is confirmed by the fact that the \diver performance is apparently better, giving the correct global minimum in 2D with more likelihood evaluations and better, though not adequate, performance in 3D. The \particlefilter algorithm works in 2D, but this is almost certainly due to the fact that it has performed a large number of likelihood evaluations in a low-dimensional space. It is outperformed by \randomsampling in 3D, as are all algorithms except \pygmoabc and \turbo. Finally, it is worth comparing the performance of the two \bo algorithms. The failure of \gpyopt to find the global minimum in any dimensionality is not unexpected, as a sharply-spiked local minimum is exactly the case that is expected to be missed by \bo due to its relatively low number of samples of the objective function, and concentration of those samples in areas where the algorithm thinks it has found interesting points. \turbo, meanwhile, is able to find the global minimum correctly in 2D (and reasonably well in 3D) because it breaks up the space into separate regions, and runs an independent \bo within each of them. One of these regions is small enough to contain the global minimum as the obvious minimum, rather than the plateau of false minima at the edge of the function range.


For the second analytic function, it is interesting to see that most algorithms are able to find the global minimum. Moreover, the fact that most of these algorithms are able to systematically outperform \randomsampling indicates that these algorithms are in fact effective methods for optimisation problems that resemble Analytic Function 2, with a large number of local minima but a single clear global minimum. \ampgo is the only algorithm that fails in all numbers of dimensions, being outperformed by \randomsampling, followed by \gpyopt, which only properly succeeds in 2D. \turbo performs better, giving adequate performance in 2D and 3D, but at the cost of a large number of total likelihood evaluations. The best algorithm is the \pygmo implementation of \de, which gets the correct answer in all dimensionalities with a low number of likelihood evaluations. \diver also performs well, with a higher number of evaluations, but the \pygmo results suggest that fewer \diver evaluations would still give good performance.  It is interesting to note that the \pygmoabc still performs well, getting the correct answer in all dimensionalities with a relatively low number of evaluations. A final interesting feature of the results in Figure~\ref{fig:result_hidden_2} is that -- as expected -- the performance of each algorithm deteriorates with increasing dimensionality, as can be seen from the increased spread of results for higher dimensionalities.

The results for Analytic Function 3 in Figure~\ref{fig:result_hidden_3} show that \ampgo retains its status as the worst algorithm, once again being outperformed by \randomsampling. However, in 2D the global minimum is almost found, with a fairly modest number of likelihood evaluations. In general, all of the algorithms now show very good performance, which is a reflection of the fact that there are now many global minima in the function, and it is easy to find at least one of them. This is further reflected in the fact that the precise configuration of each algorithm now becomes less important, with much less variation in the results of runs with different hyperparameter settings. Indeed, this is the only example among the analytic functions where a local optimisation approach would work. \gpyopt now shines, and finds the global minimum correctly in all dimensionalities with the smallest number of total likelihood evaluations. \pygmoabc still does a good job, with a relatively small number of likelihood evaluations. Comparing the \de implementations we see that \diver does not quite get the global minimum in 7D, whilst the \pygmode implementation finds it in all cases, with fewer likelihood evaluations. \turbo is able to match the performance of \gpyopt, but with more likelihood evaluations due to the requirement of running many separate optimisations (these extra optimisations are redundant in this case). 

Analytic Function 4 has various local minima, but only one global minimum. In Figure~\ref{fig:result_hidden_4}, we see that \randomsampling outperforms \ampgo again. \pygmoabc gets the answer right in all dimensionalities, as do the two \de algorithms, \cmaes, and \pso. The best performing algorithm is now the \pygmo implementation of \de, although again we should caution that \diver may give similar performance for a suitable choice of its hyperparameters. Again we see \bo fail as the dimensionality increases, although \turbo is better than \gpyopt in >5 dimensions. Of particular note is the fact that the \gwo now does not perform well at all, and seems only to have worked for Analytic Function 2 and Analytic Function 3.

In summarising the performance of the 11 different algorithms on the four different analytic functions, we can ask whether any algorithm emerges with acceptable performance on all of them.  A summary of their performance is given in Table~\ref{tab:Summary}. \pygmoabc emerges as perhaps the best candidate, since it performs well for all functions except Analytic Function 1, and it gives the best performance in that case. \ampgo is poor and consistently worse than \randomsampling. The success or failure of \bo (\gpyopt and \turbo) is interesting to investigate across the analytic functions. It generally fails when there are hidden minima, but the situation can be improved by adding latin hypercube sampling such as that found in the \turbo algorithm. \de is consistently strong in both of the \pygmo and \diver implementations, whilst \cmaes also shows consistent performance (all except for Analytic Function 1). \pso is not as consistent across the analytic functions, and where it succeeds it requires a large number of evaluations. Finally, the \particlefilter algorithm struggles in higher dimensionalities, and is highly sensitive to details of its configuration.

In Appendix~\ref{app:tables}, Table~\ref{tab:HiddenFunction1} to Table~\ref{tab:HiddenFunction4}, we show the best minimum found by each algorithm for each analytic function and dimensionality. The hyperparameter settings shown in each case are those that led to the best minimum and, where multiple different runs obtained the same result, the settings are those that needed the smallest number of total likelihood evaluations to obtain that result.

\newcolumntype{C}[1]{>{\centering}m{#1}}
\newcolumntype{R}[1]{>{\RaggedLeft}m{#1}}
\begin{sidewaystable}[htbp]
    \centering
    \begin{tabular}{|m{5.5cm} | C{3.4cm} C{3.4cm} C{3.4cm} | R{3.4cm}|}
    \hline
    & \textbf{\small Finding sharp minimum} & \textbf{\small Finding global minimum} & \textbf{\small Performance with high dimensions} & \textbf{\small Average number of evaluations} \\
    \hline
    \ampgo & {\footnotesize bad} & {\footnotesize bad} & {\footnotesize bad} & {\footnotesize low} \\
    \hline
    \cmaes & {\footnotesize very low dimensions} & {\footnotesize good} & {\footnotesize good} & {\footnotesize medium} \\
    \hline
    \diver & {\footnotesize low dimensions} & {\footnotesize good} & {\footnotesize good} & {\footnotesize high} \\
    \hline
    \particlefilter & {\footnotesize very low dimensions} & {\footnotesize low dimensions} & {\footnotesize highly configuration and function dependent} & {\footnotesize high} \\
    \hline
    \gpyopt & {\footnotesize bad} & {\footnotesize low dimensions} & {\footnotesize function dependent} & {\footnotesize low} \\
    \hline
    \pso & {\footnotesize very low dimensions} & {\footnotesize good} & {\footnotesize configuration dependent} & {\footnotesize medium} \\
    \hline
    \pygmoabc & {\footnotesize low dimensions} & {\footnotesize good} & {\footnotesize good} & {\footnotesize medium} \\
    \hline
    \pygmode & {\footnotesize bad} & {\footnotesize good} & {\footnotesize good} & {\footnotesize low} \\
    \hline
    \pygmogwo & {\footnotesize bad} & {\footnotesize bad} & {\footnotesize function dependent} & {\footnotesize medium} \\
    \hline
    \turbo & {\footnotesize low dimensions} & {\footnotesize moderate dimensions} & {\footnotesize function  dependent} & {\footnotesize high} \\
    \hline
    \randomsampling & {\footnotesize low dimensions} & {\footnotesize low dimensions} & {\footnotesize function  dependent} & {\footnotesize high} \\
    \hline
    \end{tabular}
    \caption{Summary of optimisation algorithm performance. Analytic Function 1 can be used to evaluate the performance of algorithms for finding a sharp minimum. Analytic Function 4 can be used to compare the performance of algorithms for finding a global minimum. The performance of algorithms for increasing number of dimensions and average number of evaluations was compared for all four analytic functions.}
    \label{tab:Summary}
\end{sidewaystable}
\clearpage
\subsection{Particle astrophysics test problem}

\begin{figure}
    \centering
    \includegraphics[width=0.9\textwidth]{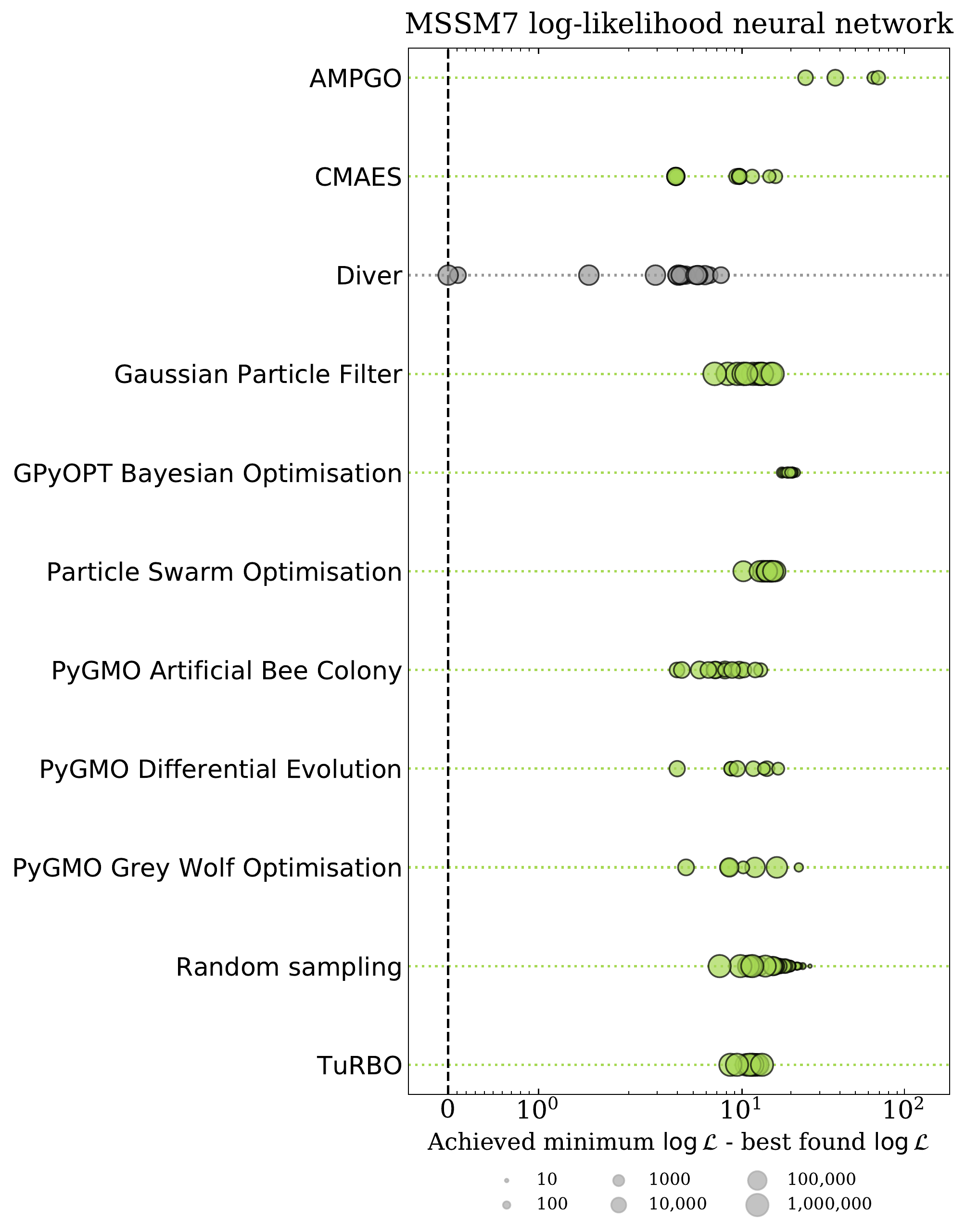}
    \caption{Results from different optimisation algorithms on the neural network approximation of the MSSM7 log-likelihood described in Section~\ref{ssec:mssm7}. The results are shown as semi-opaque circles, of which the area increases logarithmically with the number of function evaluations needed to obtain that specific result. The horizontal axis shows the difference between the log-likelihood at the found minimum and the deepest minimum found by any algorithm for any settings. To emphasise the possible bias in the test function towards \diver, the results for that algorithm are coloured differently.}
    \label{fig:result_mssm7}
\end{figure}

In Figure~\ref{fig:result_mssm7}, we show the results for each algorithm for the MSSM7 test example described in Section~\ref{ssec:mssm7}. The immediate thing to note is that \diver emerges as the best algorithm, finding the best fit of all algorithms. It comfortably outperforms the \pygmo implementations of \de, albeit with a higher number of likelihood evaluations (suggesting, once more, that the \pygmo code may give better results for different choices of the hyperparameters). The cause for this exceptionally strong performance of \diver might be the fact that the training data for the neural network was itself sampled by \diver. Although the training data was created independently of any of the optimisation experiments presented here, a neural network trained on that data might still encode the patterns typically explored by \diver, while not encoding the patterns used by the other algorithms equally well. However, the neural network still provides an example of a physically-motivated function and it is true to say that the non-\diver algorithms were not able to find the minimum of this function.

Apart from \diver, many of the algorithms give similar performance, comparing favourably to \randomsampling. Although \randomsampling is, for this physics test problem, able to come as close to the same solution as many other algorithms, it needs significantly more function evaluations to achieve this performance. The \pygmoabc algorithm is, in a surprising turn of events, not notably better than most of the other algorithms, even though it performed consistently well on the analytic functions. 

The two \bo methods, \gpyopt and \turbo, are amongst the algorithms that perform better than, or comparable to, \randomsampling, but they underperform relative to other algorithms in this group. This is likely caused by the dimensionality of the problem (12D); we already saw in the results of the analytic functions that an increase in dimensionality dragged the performance of these algorithms down strongly.

The only exception to the general trend of ``at least similar performance to \randomsampling'' is \ampgo, which remains consistently poor.

\clearpage

\section{Conclusions}\label{Conclusions}
\label{sec:conclusions}

We have performed a detailed comparison of a variety of optimisation algorithms in an attempt to find new algorithms for particle astrophysics problems. Many of these algorithms have not been used in a particle astrophysics context before, and we have examined their ability to find the correct global minimum of a range of test functions. They were also tested in their ability to correctly maximise a likelihood in a realistic particle astrophysics example based on a recent global fit of a phenomenological supersymmetry model. The algorithms we investigated were \de (using two different software implementations), \pso, the \cmaesfull, \bo (in two  different forms), \gwo, \pygmoabc, \particlefilter and \ampgo. All of the algorithms used in our comparison have publicly-available software implementations. 

For each algorithm, we characterised the hyperparameters as affecting the convergence or resolution of the optimisation, or as providing hints or improving the reliability. We then ran the different algorithms with different resolution and convergence hyperparameter settings, and compared the performance on test functions of different dimensionality, and our custom implementation of the MSSM7 likelihood function. Understandably, our main conclusion is the almost facile observation that the ``best'' algorithm depends strongly on the type of function that one wishes to optimise. However, it is possible to add some further interesting conclusions:

\begin{itemize}
    \item Algorithms that emerge as the most consistent performers when evaluated on analytic functions do not necessarily give the best performance on a realistic particle astrophysics example. This is evidenced by the fact that the \pygmoabc algorithm arguably emerged as the most consistent performer on our analytic functions. It struggled to match the performance of the \de implementation \diver on the MSSM7 likelihood function, but this might be due to a bias towards \diver in this physics inspired test case.
    \item Differential evolution (in various implementations) performed consistently well across the full barrage of tests. 
    \item \ampgo performed consistently poorly on all test examples, being outperformed by \randomsampling in most cases.
    \item \bo (in two variants, standard \gpbo, implemented in \gpyopt, and the \trustregionbo algorithm, implemented in the \turbo package) performs well for functions with many global minima, but struggles in cases with very sharply-peaked minima, or multiple global minima. Performance can be enhanced by performing separate optimisations in different latin hypercubes, at the cost of increasing the total number of likelihood evaluations. Even then, the performance degraded significantly for the analytic functions once the dimensionality increased.
\end{itemize}

Finally, many of the algorithms used here show promising performance both on the analytic functions and the particle astrophysics example. This certainly motivates their use in real-world particle astrophysics applications, and we look forward to reading future examples of their application.

\appendix
\section{Description of DarkMachines sampling framework}
\label{app:python}
All experiments run for this paper were performed in an open-source Python package written specifically for this research: the \hdsfull framework, henceforth abbreviated to \hds. This appendix outlines the general workings of, and the design ideas behind this package. It is however by no means a manual. A more complete and more technical introduction to the package can be found on the wiki at the project \github page: \url{https://github.com/DarkMachines/high-dimensional-sampling/}. The full code is published under the MIT license.

\subsubsection*{Design considerations}
As many different algorithms needed to be tested for this paper, a consistent test framework was needed. Ideally this meant that this framework would be invariant under changes of the optimisation algorithm used for investigation: it should just use the supplied optimisation method and test its performance just like it would test the performance of any other optimisation method when supplied. As the intention was to use a wide range of optimisation methods -- of which a significant number were already implemented in packages like \scannerbit~\cite{ScannerBit} -- no single existing package was found that perfectly matched our requirements.

In designing and writing the custom package the following guidelines were therefore followed:
\begin{enumerate}
    \item The package should make experiments reproducible;
    \item The evaluation of the performance of an optimisation algorithm should not depend on the exact algorithm under investigation;
    \item The package should automate as much of the experiments as possible, with a minimum loss of configurability;
    \item The package should be easy to use and install, as experiments will be performed by many people on many different machines;
    \item The output of the package should make it possible to easily compare the performance of different algorithms.
\end{enumerate}

\subsubsection*{Core components}
Following guideline (2) the package does indeed not depend on any optimisation algorithm: experiments can run with any implemented algorithm and these can be freely interchanged. However, to achieve this independence a common interface is needed. In the \hds framework this interface is implemented as the \texttt{Procedure} class. This class implements a small selection of functions that give other parts of the framework the possibility to query the procedure for new samples or for its status (e.g., whether or not it has finished sampling and whether or not it can optimise a specific target function). As the interface is very minimal (there are only 5 functions that need to be implemented), implementing new optimisation methods is relatively easy. Implemented Procedures can be found in the \texttt{optimisation} submodule of the package. As some of these require the installation of extra third-party packages, it is recommended to read the documentation on the \github wiki for more information.

Optimisation procedures are tested on test functions. To open up the possibility of having third-party test functions (i.e., functions defined by an external likelihood evaluation procedure), these functions also have an interface class: \texttt{TestFunction}. The majority of the \texttt{TestFunction}s have their analytic form programmed directly in the python code. They include common optimisation targets, like the Himmelblau function, the Rastrigin function and the ThreeHumpCamel function (see the \github wiki for a complete list of implemented functions and references to their analytic forms).

Although useful as tests of optimisation procedures, the fact that the user has access to the analytic form of these functions (or can just google for the coordinate and functional value of the optimum), they are not true blind tests of optimisation procedures. Because of this, four additional functions are implemented for which the pre-compiled binaries are included in the package. The analytic forms of these functions are not included in the package. These so-called \texttt{HiddenFunction}s are the functions referred to in Section~\ref{sec:hidden}.

Experiments are run using instances of the \texttt{Experiment} class. It is this class that guarantees the first and third design guidelines. Providing the configured optimisation \texttt{Procedure} and defining on which \texttt{TestFunction}s the procedure should be tested, the \texttt{Experiment} class runs the experiment and outputs all necessary information to interpret the procedure's performance. This information includes:

\begin{itemize}
    \item Benchmarks of the speed of the machine on which the computer is run;
    \item For each used \texttt{TestFunction}:\begin{itemize}
        \item Meta data about the Procedure and the TestFunction (e.g., values of the configurable parameters);
        \item The number of times the function was called (and if applicable the number of times the function was queried for its derivative);
        \item The coordinates of the taken samples;
        \item The found optimum and the function value at that coordinate;
    \end{itemize}
\end{itemize}

\section{Best found results and parameter settings}~\label{app:tables}
Tables~\ref{tab:HiddenFunction1} to~\ref{tab:mssm7} show for each explored optimisation algorithm the configuration and result of the run which came closest to the global minimum (for the analytic functions) or the overall best found minimum (for the MSSM7 function). If multiple runs resulted in the same function value, the result with the least number of function evaluations is shown. 

\clearpage
\begin{table}[t]
\centering
\resizebox{\textwidth}{!}{\footnotesize \begin{tabular}{|l|c|c|c|c|}
\hline
\textbf{Algorithm} & \textbf{dim} & \textbf{Parameters} & \textbf{min} & $\mathbf{N}_{\rm eval}$ \\
\hline 
\multirow{ 4}{*}{\ampgo}&2&&-0.0&5013\\
&3&&-0.0&20016\\
&5&&-0.0&10008\\
&7&&-0.0&20120\\
\hline 
\multirow{ 4}{*}{\cmaes}&2&convergence=1.0000000000000001e-11, resolution=100.0&-0.722&10900\\
&3&convergence=0.1, resolution=20.0&0.0&160\\
&5&convergence=1.0000000000000001e-11, resolution=20.0&0.0&20\\
&7&convergence=0.1, resolution=20.0&0.0&20\\
\hline 
\multirow{ 4}{*}{\diver}&2&convthresh=0.0001, np=5000&-0.998&65000\\
&3&convthresh=0.1, np=10000&-0.735&110000\\
&5&convthresh=0.001, np=2000&0.0&22000\\
&7&convthresh=0.1, np=2000&0.0&22000\\
\hline 
\multirow{ 4}{*}{\particlefilter}&2&logaritmic=True, survival\_rate=0.2, width\_decay=0.9&-1.0&225589\\
&3&logaritmic=False, survival\_rate=0.5, width\_decay=0.9&0.0&469469\\
&5&logaritmic=True, survival\_rate=0.5, width\_decay=0.95&0.0&983262\\
&7&logaritmic=True, survival\_rate=0.2, width\_decay=0.95&0.0&977481\\
\hline 
\multirow{ 4}{*}{\gpyopt}&2&eps=0.1&0.0&511\\
&3&eps=0.0001&0.0&425\\
&5&eps=0.01&0.0&345\\
&7&eps=0.001&0.0&456\\
\hline 
\multirow{ 4}{*}{\pso}&2&convthresh=0.001, np=10000&-1.0&4400\\
&3&convthresh=0.1, np=20000&0.0&4000\\
&5&convthresh=0.1, np=20000&0.0&4000\\
&7&convthresh=0.1, np=10000&0.0&4000\\
\hline 
\multirow{ 4}{*}{\pygmoabc}&2&generations=750, limit=50&-1.0&30020\\
&3&generations=750, limit=100&-0.997&30020\\
&5&generations=100, limit=50&0.0&4020\\
&7&generations=100, limit=10&0.0&4020\\
\hline 
\multirow{ 4}{*}{\pygmode}&2&generations=500, variant=iDE&0.0&80\\
&3&generations=750, variant=jDE&0.0&60\\
&5&generations=250, variant=iDE&0.0&40\\
&7&generations=750, variant=jDE&0.0&40\\
\hline 
\multirow{ 4}{*}{\pygmogwo}&2&generations=10&0.0&220\\
&3&generations=10&0.0&220\\
&5&generations=10&0.0&220\\
&7&generations=10&0.0&220\\
\hline 
\multirow{ 4}{*}{\randomsampling}&2&n\_samples=1000000&-1.0&1000000\\
&3&n\_samples=1000000&-0.903&1000000\\
&5&n\_samples=10&0.0&10\\
&7&n\_samples=10&0.0&10\\
\hline 
\multirow{ 4}{*}{\turbo}&2&max\_eval=100&-1.0&1001876\\
&3&max\_eval=100&-0.917&1052097\\
&5&max\_eval=64&0.0&650000\\
&7&max\_eval=64&0.0&650000\\
\hline 
\end{tabular}} 
\caption{Best obtained result for Analytic Function 1 (Equation~\ref{eq:hidden_1}). The `best' result is the result with the lowest found function value. If multiple samples found the same value, the result with the fewest number of needed function evaluations is shown.}
\label{tab:HiddenFunction1}
\end{table}

\begin{table}[t]
\centering
\resizebox{\textwidth}{!}{\footnotesize \begin{tabular}{|l|c|c|c|c|}
\hline
\textbf{Algorithm} & \textbf{dim} & \textbf{Parameters} & \textbf{min} & $\mathbf{N}_{\rm eval}$ \\
\hline 
\multirow{ 4}{*}{\ampgo}&2&&15.097&10062\\
&3&&45.417&20000\\
&5&&107.909&10032\\
&7&&109.605&10456\\
\hline 
\multirow{ 4}{*}{\cmaes}&2&convergence=1.0000000000000001e-11, resolution=20.0&0.0&1500\\
&3&convergence=1.0000000000000001e-11, resolution=20.0&0.0&1780\\
&5&convergence=1.0000000000000001e-11, resolution=100.0&0.0&6900\\
&7&convergence=1.0000000000000001e-11, resolution=500.0&0.0&34500\\
\hline 
\multirow{ 4}{*}{\diver}&2&convthresh=0.0001, np=10000&0.0&380000\\
&3&convthresh=0.0001, np=20000&0.0&1040000\\
&5&convthresh=0.0001, np=20000&0.0&1600000\\
&7&convthresh=0.0001, np=20000&0.0&2000000\\
\hline 
\multirow{ 4}{*}{\particlefilter}&2&logaritmic=False, survival\_rate=0.5, width\_decay=0.9&0.0&225589\\
&3&logaritmic=False, survival\_rate=0.5, width\_decay=0.95&0.0&965349\\
&5&logaritmic=False, survival\_rate=0.2, width\_decay=0.9&0.0&983262\\
&7&logaritmic=True, survival\_rate=0.5, width\_decay=0.95&3.266&977483\\
\hline 
\multirow{ 4}{*}{\gpyopt}&2&eps=0.0001&0.002&754\\
&3&eps=0.0001&1.265&693\\
&5&eps=1e-06&5.284&587\\
&7&eps=0.001&5.829&797\\
\hline 
\multirow{ 4}{*}{\pso}&2&convthresh=0.001, np=5000&0.0&21200\\
&3&convthresh=0.001, np=2000&0.0&33600\\
&5&convthresh=0.0001, np=5000&0.0&85200\\
&7&convthresh=0.0001, np=2000&0.0&133600\\
\hline 
\multirow{ 4}{*}{\pygmoabc}&2&generations=100, limit=50&0.0&4020\\
&3&generations=250, limit=100&0.0&10020\\
&5&generations=500, limit=100&0.0&20020\\
&7&generations=500, limit=50&0.0&20020\\
\hline 
\multirow{ 4}{*}{\pygmode}&2&generations=500, variant=iDE&0.0&1600\\
&3&generations=500, variant=iDE&0.0&1860\\
&5&generations=750, variant=iDE&0.0&4320\\
&7&generations=500, variant=jDE&0.0&6080\\
\hline 
\multirow{ 4}{*}{\pygmogwo}&2&generations=100&0.0&2020\\
&3&generations=1000&0.0&20020\\
&5&generations=1000&0.0&20020\\
&7&generations=1000&0.0&20020\\
\hline 
\multirow{ 4}{*}{\randomsampling}&2&n\_samples=1000000&0.008&1000000\\
&3&n\_samples=500000&0.512&500000\\
&5&n\_samples=500000&5.873&500000\\
&7&n\_samples=1000000&20.512&1000000\\
\hline 
\multirow{ 4}{*}{\turbo}&2&max\_eval=100&0.0&1000584\\
&3&max\_eval=100&0.027&1050660\\
&5&max\_eval=100&2.044&1050025\\
&7&max\_eval=100&12.101&1050007\\
\hline 
\end{tabular}} 
\caption{Best obtained result for Analytic Function 2 (Equation~\ref{eq:hidden_2}). The `best' result is the result with the lowest found function value. If multiple samples found the same value, the result with the fewest number of needed function evaluations is shown.}
\label{tab:HiddenFunction2}
\end{table}

\begin{table}[t]
\centering
\resizebox{\textwidth}{!}{\footnotesize \begin{tabular}{|l|c|c|c|c|}
\hline
\textbf{Algorithm} & \textbf{dim} & \textbf{Parameters} & \textbf{min} & $\mathbf{N}_{\rm eval}$ \\
\hline 
\multirow{ 4}{*}{\ampgo}&2&&-0.988&10014\\
&3&&-0.699&20180\\
&5&&-0.647&10008\\
&7&&-0.447&20008\\
\hline 
\multirow{ 4}{*}{\cmaes}&2&convergence=1e-07, resolution=20.0&-1.0&760\\
&3&convergence=1e-07, resolution=20.0&-1.0&920\\
&5&convergence=0.0001, resolution=20.0&-1.0&1420\\
&7&convergence=1.0000000000000001e-11, resolution=20.0&-1.0&2340\\
\hline 
\multirow{ 4}{*}{\diver}&2&convthresh=0.0001, np=10000&-1.0&210000\\
&3&convthresh=0.1, np=20000&-1.0&220000\\
&5&convthresh=0.0001, np=20000&-0.998&420000\\
&7&convthresh=0.1, np=10000&-0.984&110000\\
\hline 
\multirow{ 4}{*}{\particlefilter}&2&logaritmic=False, survival\_rate=0.2, width\_decay=0.9&-1.0&225589\\
&3&logaritmic=False, survival\_rate=0.2, width\_decay=0.9&-1.0&469460\\
&5&logaritmic=True, survival\_rate=0.2, width\_decay=0.9&-1.0&983184\\
&7&logaritmic=True, survival\_rate=0.2, width\_decay=0.9&-0.986&975284\\
\hline 
\multirow{ 4}{*}{\gpyopt}&2&eps=1e-06&-1.0&636\\
&3&eps=0.01&-1.0&623\\
&5&eps=1e-06&-1.0&635\\
&7&eps=1e-05&-1.0&675\\
\hline 
\multirow{ 4}{*}{\pso}&2&convthresh=0.01, np=2000&-1.0&4400\\
&3&convthresh=0.0001, np=10000&-1.0&4400\\
&5&convthresh=0.0001, np=2000&-0.983&4400\\
&7&convthresh=0.001, np=5000&-0.923&4400\\
\hline 
\multirow{ 4}{*}{\pygmoabc}&2&generations=100, limit=100&-1.0&4020\\
&3&generations=100, limit=50&-1.0&4020\\
&5&generations=500, limit=100&-1.0&20020\\
&7&generations=500, limit=100&-1.0&20020\\
\hline 
\multirow{ 4}{*}{\pygmode}&2&generations=750, variant=iDE&-1.0&2420\\
&3&generations=750, variant=iDE&-1.0&3140\\
&5&generations=250, variant=iDE&-1.0&5020\\
&7&generations=750, variant=jDE&-1.0&15020\\
\hline 
\multirow{ 4}{*}{\pygmogwo}&2&generations=1000&-1.0&20020\\
&3&generations=100&-1.0&2020\\
&5&generations=1000&-1.0&20020\\
&7&generations=1000&-1.0&20020\\
\hline 
\multirow{ 4}{*}{\randomsampling}&2&n\_samples=50000&-1.0&50000\\
&3&n\_samples=1000000&-1.0&1000000\\
&5&n\_samples=1000000&-0.991&1000000\\
&7&n\_samples=1000000&-0.964&1000000\\
\hline 
\multirow{ 4}{*}{\turbo}&2&max\_eval=64&-1.0&700000\\
&3&max\_eval=100&-1.0&1050981\\
&5&max\_eval=100&-1.0&1050025\\
&7&max\_eval=100&-0.998&1050000\\
\hline 
\end{tabular}} 
\caption{Best obtained result for Analytic Function 3 (Equation~\ref{eq:hidden_3}). The `best' result is the result with the lowest found function value. If multiple samples found the same value, the result with the fewest number of needed function evaluations is shown.}
\label{tab:HiddenFunction3}
\end{table}

\begin{table}[t]
\centering
\resizebox{\textwidth}{!}{\footnotesize \begin{tabular}{|l|c|c|c|c|}
\hline
\textbf{Algorithm} & \textbf{dim} & \textbf{Parameters} & \textbf{min} & $\mathbf{N}_{\rm eval}$ \\
\hline 
\multirow{ 4}{*}{\ampgo}&2&&302.982&20004\\
&3&&576.415&20036\\
&5&&1063.57&10086\\
&7&&1874.14&20000\\
\hline 
\multirow{ 4}{*}{\cmaes}&2&convergence=0.0001, resolution=20.0&-0.0&1960\\
&3&convergence=1e-07, resolution=20.0&-0.0&3160\\
&5&convergence=1.0000000000000001e-11, resolution=20.0&-0.0&6800\\
&7&convergence=1.0000000000000001e-11, resolution=20.0&-0.0&4420\\
\hline 
\multirow{ 4}{*}{\diver}&2&convthresh=0.0001, np=2000&-0.0&64000\\
&3&convthresh=0.0001, np=2000&-0.0&88000\\
&5&convthresh=0.0001, np=5000&-0.0&315000\\
&7&convthresh=0.0001, np=5000&-0.0&365000\\
\hline 
\multirow{ 4}{*}{\particlefilter}&2&logaritmic=True, survival\_rate=0.2, width\_decay=0.95&-0.0&463869\\
&3&logaritmic=True, survival\_rate=0.2, width\_decay=0.95&-0.0&965349\\
&5&logaritmic=False, survival\_rate=0.5, width\_decay=0.99&118.806&983262\\
&7&logaritmic=True, survival\_rate=0.2, width\_decay=0.9&493.412&977483\\
\hline 
\multirow{ 4}{*}{\gpyopt}&2&eps=0.0001&0.017&651\\
&3&eps=0.01&0.062&852\\
&5&eps=1e-05&53.532&954\\
&7&eps=0.0001&928.87&1011\\
\hline 
\multirow{ 4}{*}{\pso}&2&convthresh=0.001, np=2000&-0.0&81600\\
&3&convthresh=0.001, np=20000&-0.0&110000\\
&5&convthresh=0.001, np=2000&-0.0&200400\\
&7&convthresh=0.0001, np=5000&-0.0&330000\\
\hline 
\multirow{ 4}{*}{\pygmoabc}&2&generations=100, limit=100&-0.0&4020\\
&3&generations=250, limit=100&-0.0&10020\\
&5&generations=500, limit=100&-0.0&20020\\
&7&generations=750, limit=100&-0.0&30020\\
\hline 
\multirow{ 4}{*}{\pygmode}&2&generations=100, variant=jDE&-0.0&2020\\
&3&generations=250, variant=jDE&-0.0&5020\\
&5&generations=250, variant=iDE&-0.0&5020\\
&7&generations=500, variant=jDE&-0.0&10020\\
\hline 
\multirow{ 4}{*}{\pygmogwo}&2&generations=1000&118.439&20020\\
&3&generations=1000&236.878&20020\\
&5&generations=1000&355.32&20020\\
&7&generations=50&1034.88&1020\\
\hline 
\multirow{ 4}{*}{\randomsampling}&2&n\_samples=500000&0.008&500000\\
&3&n\_samples=500000&2.682&500000\\
&5&n\_samples=1000000&184.329&1000000\\
&7&n\_samples=500000&439.961&500000\\
\hline 
\multirow{ 4}{*}{\turbo}&2&max\_eval=100&0.0&1000636\\
&3&max\_eval=100&0.034&1050318\\
&5&max\_eval=100&0.395&1050010\\
&7&max\_eval=100&178.766&1050000\\
\hline 
\end{tabular}} 
\caption{Best obtained result for Analytic Function 4 (Equation~\ref{eq:hidden_4}). The `best' result is the result with the lowest found function value. If multiple samples found the same value, the result with the fewest number of needed function evaluations is shown.}
\label{tab:HiddenFunction4}
\end{table}

\begin{table}[t]
\centering
\resizebox{\textwidth}{!}{\footnotesize \begin{tabular}{|l|c|c|c|}
\hline
\textbf{Algorithm} & \textbf{Parameters} & \textbf{min} & $\mathbf{N}_{\rm eval}$ \\
\hline 
\ampgo&&262.815&10010\\
\hline 
\cmaes&convergence=500.0, resolution=0.0001&242.121&41000\\
\hline 
\diver&convthresh=0.0001, np=20000&238.214&200000\\
\hline 
\particlefilter&logaritmic=True, survival\_rate=0.5, width\_decay=0.9&244.993&1333119\\
\hline 
\gpyopt&eps=0.0001&255.827&684\\
\hline 
\pso&convthresh=0.001, np=20000&248.399&262800\\
\hline 
\pygmoabc&generations=250, limit=50&242.189&10020\\
\hline 
\pygmode&generations=750, variant=2&242.197&15020\\
\hline 
\pygmogwo&generations=1000&242.731&20020\\
\hline 
\randomsampling&n\_samples=1000000.0&245.489&1000000\\
\hline 
\turbo&max\_evals=100&246.688&1000000\\
\hline 
\end{tabular}} 
\caption{Best obtained result for the approximation of the 12-dimensional MSSM7 log-likelihood described in Section~\ref{ssec:mssm7}. The `best' result is the result with the lowest found function value. If multiple samples found the same value, the result with the fewest number of needed function evaluations is shown.}
\label{tab:mssm7}
\end{table}

\clearpage
\section{Acknowledgements}
MvB acknowledges support from the Science and Technology Facilities Council (grant number ST/T000864/1).
R. RdA acknowledges partial funding/support from the Elusives ITN (Marie Sk\l{}odowska-Curie grant agreement No 674896), the ``SOM Sabor y origen de la Materia" (FPA 2017-85985-P). BS, LH and SC acknowledges the support by the Netherlands eScience Center under the project \href{https://www.esciencecenter.nl/projects/idark/}{iDark:The intelligent Dark Matter Survey}. AF is supported by an NSFC Research Fund for International Young Scientists grant 11950410509. MJW and CB are funded by the Australian Research Council (ARC) Discovery Project DP180102209, MJW and AL by the ARC Centre of Excellence for Dark Matter Particle Physics CE200100008, and PS by ARC Future Fellowship FT190100814. JM acknowledges the support from the grant PGC2018-094856-B-I00 by the Spanish MICIU / AEI and the European Union / FEDER, and the APOSTD/2019/165 fellowship by the University of Valencia, Generalitat Valenciana and European Union. EGM gratefully
acknowledge the use of the facilities of Centro de Computaci\'on Cient\'ifica (CCC) at Universidad Aut\'onoma de Madrid. EGM also acknowledge financial support from Spanish Plan Nacional I+D+i, grants TIN2016-76406-P and PID2019-106827GB-I00 / AEI / 10.13039/501100011033. 

\bibliographystyle{JHEP}
\bibliography{HighDimensionalSampling}

\providecommand{\href}[2]{#2}\begingroup\raggedright\begin{thebibliography}{100}

\bibitem{AbdusSalam:2020rdj}
{\bf \GB, \textsf{MasterCode}, \textsf{Fittino}, \textsf{HEPfit} and
  \textsf{BayesFits}} Collaboration, S.~S. AbdusSalam et~al., {\it {Simple and
  statistically sound strategies for analysing physical theories}},
  \href{http://arxiv.org/abs/2012.09874}{{\tt arXiv:2012.09874}}.

\bibitem{Cousins:2020ntk}
R.~D. Cousins, {\it {What is the likelihood function, and how is it used in
  particle physics?}},  \href{http://arxiv.org/abs/2010.00356}{{\tt
  arXiv:2010.00356}}.

\bibitem{Cowan:2010js}
G.~Cowan, K.~Cranmer, E.~Gross, and O.~Vitells, {\it {Asymptotic formulae for
  likelihood-based tests of new physics}},  {\em Eur. Phys. J. C} {\bf 71}
  (2011) 1554, [\href{http://arxiv.org/abs/1007.1727}{{\tt arXiv:1007.1727}}].
  [Erratum: Eur.Phys.J.C 73, 2501 (2013)].

\bibitem{Balazs:2017moi}
{\bf \GB} Collaboration, C.~Bal\'azs et~al., {\it {ColliderBit: a GAMBIT module
  for the calculation of high-energy collider observables and likelihoods}},
  {\em Eur. Phys. J. C} {\bf 77} (2017), no.~11 795,
  [\href{http://arxiv.org/abs/1705.07919}{{\tt arXiv:1705.07919}}].

\bibitem{RiosReview}
L.~Rios and N.~Sahinidis, {\it Derivative-free optimization: A review of
  algorithms and comparison of software implementations},  {\em Journal of
  Global Optimization} {\bf 56} (11, 2009).

\bibitem{James:1975dr}
F.~James and M.~Roos, {\it {Minuit: A System for Function Minimization and
  Analysis of the Parameter Errors and Correlations}},  {\em Comput. Phys.
  Commun.} {\bf 10} (1975) 343--367.

\bibitem{Feroz:2011bj}
F.~Feroz, K.~Cranmer, M.~Hobson, R.~Ruiz~de Austri, and R.~Trotta, {\it
  {Challenges of Profile Likelihood Evaluation in Multi-Dimensional SUSY
  Scans}},  {\em JHEP} {\bf 06} (2011) 042,
  [\href{http://arxiv.org/abs/1101.3296}{{\tt arXiv:1101.3296}}].

\bibitem{Baltz04}
E.~A. {Baltz} and P.~{Gondolo}, {\it {Markov Chain Monte Carlo Exploration of
  Minimal Supergravity with Implications for Dark Matter}},  {\em \jhep} {\bf
  10} (2004) 52, [\href{http://arxiv.org/abs/hep-ph/0407039}{{\tt
  hep-ph/0407039}}].

\bibitem{Allanach06}
B.~C. {Allanach} and C.~G. {Lester}, {\it {Multidimensional mSUGRA likelihood
  maps}},  {\em \prd} {\bf 73} (2006), no.~1 015013,
  [\href{http://arxiv.org/abs/hep-ph/0507283}{{\tt hep-ph/0507283}}].

\bibitem{SFitter}
R.~{Lafaye}, T.~{Plehn}, and D.~{Zerwas}, {\it {SFITTER: SUSY Parameter
  Analysis at LHC and LC}},  \href{http://arxiv.org/abs/hep-ph/0404282}{{\tt
  hep-ph/0404282}}.

\bibitem{Ruiz06}
R.~{Ruiz de Austri}, R.~{Trotta}, and L.~{Roszkowski}, {\it {A Markov chain
  Monte Carlo analysis of CMSSM}},  {\em \jhep} {\bf 5} (2006) 2,
  [\href{http://arxiv.org/abs/hep-ph/0602028}{{\tt hep-ph/0602028}}].

\bibitem{Strege15}
C.~{Strege}, G.~{Bertone}, G.~J. {Besjes}, S.~{Caron}, R.~{Ruiz de Austri},
  A.~{Strubig}, and R.~{Trotta}, {\it {Profile likelihood maps of a
  15-dimensional MSSM}},  {\em \jhep} {\bf 9} (Sept., 2014) 81,
  [\href{http://arxiv.org/abs/1405.0622}{{\tt arXiv:1405.0622}}].

\bibitem{Fittinocoverage}
P.~{Bechtle}, J.~E. {Camargo-Molina}, K.~{Desch}, H.~K. {Dreiner}, M.~{Hamer},
  M.~{Kr{\"a}mer}, B.~{O'Leary}, W.~{Porod}, B.~{Sarrazin}, T.~{Stefaniak},
  M.~{Uhlenbrock}, and P.~{Wienemann}, {\it {Killing the cMSSM softly}},  {\em
  \epjc} {\bf 76} (Feb., 2016) 96, [\href{http://arxiv.org/abs/1508.05951}{{\tt
  arXiv:1508.05951}}].

\bibitem{Catalan:2015cna}
M.~E. Cabrera-Catalan, S.~Ando, C.~Weniger, and F.~Zandanel, {\it {Indirect and
  direct detection prospect for TeV dark matter in the nine parameter MSSM}},
  {\em \prd} {\bf 92} (2015), no.~3 035018,
  [\href{http://arxiv.org/abs/1503.00599}{{\tt arXiv:1503.00599}}].

\bibitem{MasterCodeMSSM10}
K.~J. {de Vries}, E.~A. {Bagnaschi}, O.~{Buchmueller}, R.~{Cavanaugh},
  M.~{Citron}, A.~{De Roeck}, M.~J. {Dolan}, J.~R. {Ellis}, H.~{Fl{\"a}cher},
  S.~{Heinemeyer}, G.~{Isidori}, S.~{Malik}, J.~{Marrouche}, D.~M. {Santos},
  K.~A. {Olive}, K.~{Sakurai}, and G.~{Weiglein}, {\it {The pMSSM10 after LHC
  run 1}},  {\em \epjc} {\bf 75} (Sept., 2015) 422,
  [\href{http://arxiv.org/abs/1504.03260}{{\tt arXiv:1504.03260}}].

\bibitem{2007NewAR..51..316T}
R.~{Trotta}, R.~R. {de Austri}, and L.~{Roszkowski}, {\it {Prospects for direct
  dark matter detection in the constrained MSSM}},  {\em New Astronomy Review}
  {\bf 51} (2007) 316--320, [\href{http://arxiv.org/abs/astro-ph/0609126}{{\tt
  astro-ph/0609126}}].

\bibitem{2007JHEP...07..075R}
L.~{Roszkowski}, R.~{Ruiz de Austri}, and R.~{Trotta}, {\it {Implications for
  the Constrained MSSM from a new prediction for b {$\to$} s{$\gamma$}}},  {\em
  \jhep} {\bf 7} (2007) 75, [\href{http://arxiv.org/abs/0705.2012}{{\tt
  0705.2012}}].

\bibitem{Roszkowski09a}
L.~{Roszkowski}, R.~{Ruiz de Austri}, J.~{Silk}, and R.~{Trotta}, {\it {On
  prospects for dark matter indirect detection in the Constrained MSSM}},  {\em
  \plb} {\bf 671} (2009) 10--14, [\href{http://arxiv.org/abs/0707.0622}{{\tt
  0707.0622}}].

\bibitem{Martinez09}
G.~D. {Martinez}, J.~S. {Bullock}, M.~{Kaplinghat}, L.~E. {Strigari}, and
  R.~{Trotta}, {\it {Indirect Dark Matter detection from Dwarf satellites:
  joint expectations from astrophysics and supersymmetry}},  {\em \jcap} {\bf
  6} (2009) 14, [\href{http://arxiv.org/abs/0902.4715}{{\tt 0902.4715}}].

\bibitem{Roszkowski09b}
L.~{Roszkowski}, R.~{Ruiz de Austri}, R.~{Trotta}, Y.-L.~S. {Tsai}, and T.~A.
  {Varley}, {\it {Global fits of the nonuniversal Higgs model}},  {\em \prd}
  {\bf 83} (Jan., 2011) 015014, [\href{http://arxiv.org/abs/0903.1279}{{\tt
  arXiv:0903.1279}}].

\bibitem{Roszkowski10}
L.~{Roszkowski}, R.~{Ruiz de Austri}, and R.~{Trotta}, {\it {Efficient
  reconstruction of constrained MSSM parameters from LHC data: A case study}},
  {\em \prd} {\bf 82} (Sept., 2010) 055003,
  [\href{http://arxiv.org/abs/0907.0594}{{\tt arXiv:0907.0594}}].

\bibitem{Scott09c}
P.~{Scott}, J.~{Conrad}, J.~{Edsj{\"o}}, L.~{Bergstr{\"o}m}, C.~{Farnier}, and
  Y.~{Akrami}, {\it {Direct constraints on minimal supersymmetry from Fermi-LAT
  observations of the dwarf galaxy Segue 1}},  {\em \jcap} {\bf 1} (Jan., 2010)
  31, [\href{http://arxiv.org/abs/0909.3300}{{\tt arXiv:0909.3300}}].

\bibitem{Akrami09}
Y.~{Akrami}, P.~{Scott}, J.~Edsj{\"o}, J.~{Conrad}, and L.~{Bergstr{\"o}m},
  {\it {A profile likelihood analysis of the Constrained MSSM with genetic
  algorithms}},  {\em \jhep} {\bf 4} (2010) 57,
  [\href{http://arxiv.org/abs/0910.3950}{{\tt arXiv:0910.3950}}].

\bibitem{BertoneLHCDD}
G.~{Bertone}, D.~G. {Cerde{\~n}o}, M.~{Fornasa}, R.~{Ruiz de Austri}, and
  R.~{Trotta}, {\it {Identification of dark matter particles with LHC and
  direct detection data}},  {\em \prd} {\bf 82} (Sept., 2010) 055008,
  [\href{http://arxiv.org/abs/1005.4280}{{\tt arXiv:1005.4280}}].

\bibitem{Akrami:2010dn}
Y.~Akrami, C.~Savage, P.~Scott, J.~Conrad, and J.~Edsj{\"o}, {\it {How well
  will ton-scale dark matter direct detection experiments constrain minimal
  supersymmetry?}},  {\em \jcap} {\bf 1104} (2011) 012,
  [\href{http://arxiv.org/abs/1011.4318}{{\tt arXiv:1011.4318}}].

\bibitem{SBCoverage}
M.~{Bridges}, K.~{Cranmer}, F.~{Feroz}, M.~{Hobson}, R.~{Ruiz de Austri}, and
  R.~{Trotta}, {\it {A coverage study of CMSSM based on ATLAS sensitivity using
  fast neural networks techniques}},  {\em \jhep} {\bf 3} (Mar., 2011) 12,
  [\href{http://arxiv.org/abs/1011.4306}{{\tt arXiv:1011.4306}}].

\bibitem{Nightmare}
G.~{Bertone}, D.~{Cumberbatch}, R.~{Ruiz de Austri}, and R.~{Trotta}, {\it
  {Dark Matter searches: the nightmare scenario}},  {\em \jcap} {\bf 1} (Jan.,
  2012) 4, [\href{http://arxiv.org/abs/1107.5813}{{\tt arXiv:1107.5813}}].

\bibitem{BertoneLHCID}
G.~{Bertone}, D.~G. {Cerde{\~n}o}, M.~{Fornasa}, L.~{Pieri}, R.~{Ruiz de
  Austri}, and R.~{Trotta}, {\it {Complementarity of indirect and accelerator
  dark matter searches}},  {\em \prd} {\bf 85} (Mar., 2012) 055014,
  [\href{http://arxiv.org/abs/1111.2607}{{\tt arXiv:1111.2607}}].

\bibitem{Akrami11coverage}
Y.~{Akrami}, C.~{Savage}, P.~{Scott}, J.~{Conrad}, and J.~{Edsj{\"o}}, {\it
  {Statistical coverage for supersymmetric parameter estimation: a case study
  with direct detection of dark matter}},  {\em \jcap} {\bf 7} (July, 2011) 2,
  [\href{http://arxiv.org/abs/1011.4297}{{\tt arXiv:1011.4297}}].

\bibitem{IC22Methods}
P.~{Scott}, C.~{Savage}, J.~{Edsj{\"o}}, and {the IceCube Collaboration:
  R.~Abbasi et al.}, {\it {Use of event-level neutrino telescope data in global
  fits for theories of new physics}},  {\em \jcap} {\bf 11} (Nov., 2012) 57,
  [\href{http://arxiv.org/abs/1207.0810}{{\tt arXiv:1207.0810}}].

\bibitem{SuperbayesXENON100}
G.~{Bertone}, D.~G. {Cerde{\~n}o}, M.~{Fornasa}, R.~{Ruiz de Austri},
  C.~{Strege}, and R.~{Trotta}, {\it {Global fits of the cMSSM including the
  first LHC and XENON100 data}},  {\em \jcap} {\bf 1} (Jan., 2012) 15,
  [\href{http://arxiv.org/abs/1107.1715}{{\tt arXiv:1107.1715}}].

\bibitem{SuperBayesGC}
G.~{Bertone}, F.~{Calore}, S.~{Caron}, R.~{Ruiz}, J.~S. {Kim}, R.~{Trotta}, and
  C.~{Weniger}, {\it {Global analysis of the pMSSM in light of the Fermi GeV
  excess: prospects for the LHC Run-II and astroparticle experiments}},  {\em
  \jcap} {\bf 2016} (Apr, 2016) 037,
  [\href{http://arxiv.org/abs/1507.07008}{{\tt arXiv:1507.07008}}].

\bibitem{Buchmueller08}
O.~{Buchmueller}, R.~{Cavanaugh}, A.~{DeRoeck}, J.~R. {Ellis},
  H.~{Fl{\"a}cher}, S.~{Heinemeyer}, G.~{Isidori}, K.~A. {Olive},
  P.~{Paradisi}, F.~J. {Ronga}, and G.~{Weiglein}, {\it {Predictions for
  supersymmetric particle masses using indirect experimental and cosmological
  constraints}},  {\em \jhep} {\bf 9} (2008) 117,
  [\href{http://arxiv.org/abs/0808.4128}{{\tt arXiv:0808.4128}}].

\bibitem{Buchmueller09}
O.~{Buchmueller}, R.~{Cavanaugh}, A.~{de Roeck}, J.~R. {Ellis}, H.~{Flaecher},
  S.~{Heinemeyer}, G.~{Isidori}, K.~A. {Olive}, F.~J. {Ronga}, and
  G.~{Weiglein}, {\it {Likelihood functions for supersymmetric observables in
  frequentist analyses of the CMSSM and NUHM1}},  {\em \epjc} {\bf 64} (Dec.,
  2009) 391--415, [\href{http://arxiv.org/abs/0907.5568}{{\tt
  arXiv:0907.5568}}].

\bibitem{MasterCodemSUGRA}
O.~{Buchmueller}, R.~{Cavanaugh}, D.~{Colling}, A.~{de Roeck}, M.~J. {Dolan},
  J.~R. {Ellis}, H.~{Fl{\"a}cher}, S.~{Heinemeyer}, K.~A. {Olive},
  S.~{Rogerson}, F.~J. {Ronga}, and G.~{Weiglein}, {\it {Frequentist analysis
  of the parameter space of minimal supergravity}},  {\em \epjc} {\bf 71}
  (Mar., 2011) 1583, [\href{http://arxiv.org/abs/1011.6118}{{\tt
  arXiv:1011.6118}}].

\bibitem{MasterCode11}
O.~{Buchmueller}, R.~{Cavanaugh}, D.~{Colling}, A.~{de Roeck}, M.~J. {Dolan},
  J.~R. {Ellis}, H.~{Fl{\"a}cher}, S.~{Heinemeyer}, G.~{Isidori}, K.~{Olive},
  S.~{Rogerson}, F.~{Ronga}, and G.~{Weiglein}, {\it {Implications of initial
  LHC searches for supersymmetry}},  {\em \epjc} {\bf 71} (May, 2011) 1634,
  [\href{http://arxiv.org/abs/1102.4585}{{\tt arXiv:1102.4585}}].

\bibitem{MastercodeXENON100}
O.~{Buchmueller}, R.~{Cavanaugh}, D.~{Colling}, A.~{De Roeck}, M.~J. {Dolan},
  J.~R. {Ellis}, H.~{Fl{\"a}cher}, S.~{Heinemeyer}, G.~{Isidori},
  D.~{Mart{\'{\i}}nez Santos}, K.~A. {Olive}, S.~{Rogerson}, F.~J. {Ronga}, and
  G.~{Weiglein}, {\it {Supersymmetry and dark matter in light of LHC 2010 and
  XENON100 data}},  {\em \epjc} {\bf 71} (Aug., 2011) 1722,
  [\href{http://arxiv.org/abs/1106.2529}{{\tt arXiv:1106.2529}}].

\bibitem{MastercodeHiggs}
O.~{Buchmueller}, R.~{Cavanaugh}, A.~{De Roeck}, M.~J. {Dolan}, J.~R. {Ellis},
  H.~{Fl{\"a}cher}, S.~{Heinemeyer}, G.~{Isidori}, J.~{Marrouche},
  D.~{Mart{\'{\i}}nez Santos}, K.~A. {Olive}, S.~{Rogerson}, F.~J. {Ronga},
  K.~J. {de Vries}, and G.~{Weiglein}, {\it {Higgs and supersymmetry}},  {\em
  \epjc} {\bf 72} (June, 2012) 2020,
  [\href{http://arxiv.org/abs/1112.3564}{{\tt arXiv:1112.3564}}].

\bibitem{Buchmueller:2014yva}
O.~Buchmueller et~al., {\it {The NUHM2 after LHC Run 1}},  {\em \epjc} {\bf 74}
  (2014), no.~12 3212, [\href{http://arxiv.org/abs/1408.4060}{{\tt
  arXiv:1408.4060}}].

\bibitem{Bagnaschi:2016afc}
E.~Bagnaschi et~al., {\it {Likelihood Analysis of Supersymmetric SU(5) GUTs}},
  {\em \epjc} {\bf 77} (2017), no.~2 104,
  [\href{http://arxiv.org/abs/1610.10084}{{\tt arXiv:1610.10084}}].

\bibitem{Bagnaschi:2016xfg}
E.~Bagnaschi et~al., {\it {Likelihood Analysis of the Minimal AMSB Model}},
  {\em \epjc} {\bf 77} (2017), no.~4 268,
  [\href{http://arxiv.org/abs/1612.05210}{{\tt arXiv:1612.05210}}].

\bibitem{Allanach:2007qk}
B.~C. Allanach, K.~Cranmer, C.~G. Lester, and A.~M. Weber, {\it {Natural
  priors, CMSSM fits and LHC weather forecasts}},  {\em \jhep} {\bf 08} (2007)
  023, [\href{http://arxiv.org/abs/0705.0487}{{\tt arXiv:0705.0487}}].

\bibitem{Abdussalam09a}
S.~S. {Abdussalam}, B.~C. {Allanach}, F.~{Quevedo}, F.~{Feroz}, and
  M.~{Hobson}, {\it {Fitting the phenomenological MSSM}},  {\em \prd} {\bf 81}
  (May, 2010) 095012, [\href{http://arxiv.org/abs/0904.2548}{{\tt
  arXiv:0904.2548}}].

\bibitem{Abdussalam09b}
S.~S. {Abdussalam}, B.~C. {Allanach}, M.~J. {Dolan}, F.~{Feroz}, and M.~P.
  {Hobson}, {\it {Selecting a model of supersymmetry breaking mediation}},
  {\em \prd} {\bf 80} (Aug., 2009) 035017,
  [\href{http://arxiv.org/abs/0906.0957}{{\tt arXiv:0906.0957}}].

\bibitem{Allanach11b}
B.~C. {Allanach}, {\it {Impact of CMS Multi-jets and Missing Energy Search on
  CMSSM Fits}},  {\em \prd} {\bf 83} (May, 2011) 095019,
  [\href{http://arxiv.org/abs/1102.3149}{{\tt arXiv:1102.3149}}].

\bibitem{Allanach11a}
B.~C. {Allanach}, T.~J. {Khoo}, C.~G. {Lester}, and S.~L. {Williams}, {\it {The
  impact of ATLAS zero-lepton, jets and missing momentum search on a CMSSM
  fit}},  {\em \jhep} {\bf 6} (June, 2011) 35,
  [\href{http://arxiv.org/abs/1103.0969}{{\tt arXiv:1103.0969}}].

\bibitem{Farmer13}
C.~{Bal{\'a}zs}, A.~{Buckley}, D.~{Carter}, B.~{Farmer}, and M.~{White}, {\it
  {Should we still believe in constrained supersymmetry?}},  {\em \epjc} {\bf
  73} (Oct., 2013) 2563, [\href{http://arxiv.org/abs/1205.1568}{{\tt
  arXiv:1205.1568}}].

\bibitem{arXiv:1212.4821}
M.~E. Cabrera, J.~A. Casas, and R.~Ruiz~de Austri, {\it {The health of SUSY
  after the Higgs discovery and the XENON100 data}},  {\em \jhep} {\bf 07}
  (2013) 182, [\href{http://arxiv.org/abs/1212.4821}{{\tt arXiv:1212.4821}}].

\bibitem{Fowlie13}
A.~{Fowlie}, K.~{Kowalska}, L.~{Roszkowski}, E.~M. {Sessolo}, and Y.-L.~S.
  {Tsai}, {\it {Dark matter and collider signatures of the MSSM}},  {\em \prd}
  {\bf 88} (Sept., 2013) 055012, [\href{http://arxiv.org/abs/1306.1567}{{\tt
  arXiv:1306.1567}}].

\bibitem{Henrot14}
S.~{Henrot-Versill{\'e}}, R.~{Lafaye}, T.~{Plehn}, M.~{Rauch}, D.~{Zerwas},
  S.~{Plaszczynski}, B.~R. {d'Orfeuil}, and M.~{Spinelli}, {\it {Constraining
  supersymmetry using the relic density and the Higgs boson}},  {\em \prd} {\bf
  89} (Mar., 2014) 055017, [\href{http://arxiv.org/abs/1309.6958}{{\tt
  arXiv:1309.6958}}].

\bibitem{Kim:2013uxa}
D.~Kim, P.~Athron, C.~Bal{\'a}zs, B.~Farmer, and E.~Hutchison, {\it {Bayesian
  naturalness of the CMSSM and CNMSSM}},  {\em \prd} {\bf 90} (2014), no.~5
  055008, [\href{http://arxiv.org/abs/1312.4150}{{\tt arXiv:1312.4150}}].

\bibitem{arXiv:1503.08219}
K.~Kowalska, L.~Roszkowski, E.~M. Sessolo, and A.~J. Williams, {\it
  {GUT-inspired SUSY and the muon $g-2$ anomaly: prospects for LHC 14 TeV}},
  {\em \jhep} {\bf 06} (2015) 020, [\href{http://arxiv.org/abs/1503.08219}{{\tt
  arXiv:1503.08219}}].

\bibitem{arXiv:1604.02102}
M.~E. Cabrera, J.~A. Casas, A.~Delgado, S.~Robles, and R.~Ruiz~de Austri, {\it
  {Naturalness of MSSM dark matter}},  {\em \jhep} {\bf 08} (2016) 058,
  [\href{http://arxiv.org/abs/1604.02102}{{\tt arXiv:1604.02102}}].

\bibitem{Han:2016gvr}
C.~{Han}, K.-i. {Hikasa}, L.~{Wu}, J.~M. {Yang}, and Y.~{Zhang}, {\it {Status
  of CMSSM in light of current LHC Run-2 and LUX data}},  {\em \plb} {\bf 769}
  (Jun, 2017) 470--476, [\href{http://arxiv.org/abs/1612.02296}{{\tt
  arXiv:1612.02296}}].

\bibitem{Bechtle:2014yna}
P.~Bechtle et~al., {\it {How alive is constrained SUSY really?}},  in {\em
  {Proceedings, 37th International Conference on High Energy Physics (ICHEP
  2014): Valencia, Spain, July 2-9, 2014}}, 2016.
\newblock \href{http://arxiv.org/abs/1410.6035}{{\tt arXiv:1410.6035}}.

\bibitem{arXiv:1405.4289}
L.~Roszkowski, E.~M. Sessolo, and A.~J. Williams, {\it {What next for the CMSSM
  and the NUHM: Improved prospects for superpartner and dark matter
  detection}},  {\em \jhep} {\bf 08} (2014) 067,
  [\href{http://arxiv.org/abs/1405.4289}{{\tt arXiv:1405.4289}}].

\bibitem{arXiv:1402.5419}
A.~Fowlie and M.~Raidal, {\it {Prospects for constrained supersymmetry at
  $\sqrt{s}={33}\,\text {TeV} $ and $\sqrt{s}={100}\,\text {TeV} $
  proton-proton super-colliders}},  {\em \epjc} {\bf 74} (2014) 2948,
  [\href{http://arxiv.org/abs/1402.5419}{{\tt arXiv:1402.5419}}].

\bibitem{MastercodeCMSSM}
O.~Buchmueller et~al., {\it {The CMSSM and NUHM1 after LHC Run 1}},  {\em
  \epjc} {\bf 74} (2014), no.~6 2922,
  [\href{http://arxiv.org/abs/1312.5250}{{\tt arXiv:1312.5250}}].

\bibitem{arXiv:1312.5233}
O.~Buchmueller et~al., {\it {Implications of Improved Higgs Mass Calculations
  for Supersymmetric Models}},  {\em \epjc} {\bf 74} (2014), no.~3 2809,
  [\href{http://arxiv.org/abs/1312.5233}{{\tt arXiv:1312.5233}}].

\bibitem{arXiv:1310.3045}
P.~Bechtle et~al., {\it {Constrained Supersymmetry after the Higgs Boson
  Discovery: A global analysis with Fittino}},  {\em \pos} {\bf EPS-HEP2013}
  (2013) 313, [\href{http://arxiv.org/abs/1310.3045}{{\tt arXiv:1310.3045}}].

\bibitem{arXiv:1309.6958}
S.~Henrot-Versill\'e, R.~Lafaye, T.~Plehn, M.~Rauch, D.~Zerwas,
  S.~Plaszczynski, B.~Rouill\'e~d'Orfeuil, and M.~Spinelli, {\it {Constraining
  Supersymmetry using the relic density and the Higgs boson}},  {\em \prd} {\bf
  89} (2014), no.~5 055017, [\href{http://arxiv.org/abs/1309.6958}{{\tt
  arXiv:1309.6958}}].

\bibitem{arXiv:1307.3383}
N.~Bornhauser and M.~Drees, {\it {Determination of the CMSSM Parameters using
  Neural Networks}},  {\em \prd} {\bf 88} (2013) 075016,
  [\href{http://arxiv.org/abs/1307.3383}{{\tt arXiv:1307.3383}}].

\bibitem{arXiv:1304.5526}
S.~Akula and P.~Nath, {\it {Gluino-driven radiative breaking, Higgs boson mass,
  muon g-2, and the Higgs diphoton decay in supergravity unification}},  {\em
  \prd} {\bf 87} (2013), no.~11 115022,
  [\href{http://arxiv.org/abs/1304.5526}{{\tt arXiv:1304.5526}}].

\bibitem{arXiv:1212.2886}
M.~Citron, J.~Ellis, F.~Luo, J.~Marrouche, K.~A. Olive, and K.~J. de~Vries,
  {\it {End of the CMSSM coannihilation strip is nigh}},  {\em \prd} {\bf 87}
  (2013), no.~3 036012, [\href{http://arxiv.org/abs/1212.2886}{{\tt
  arXiv:1212.2886}}].

\bibitem{Strege13}
C.~{Strege}, G.~{Bertone}, F.~{Feroz}, M.~{Fornasa}, R.~{Ruiz de Austri}, and
  R.~{Trotta}, {\it {Global fits of the cMSSM and NUHM including the LHC Higgs
  discovery and new XENON100 constraints}},  {\em \jcap} {\bf 4} (Apr., 2013)
  13, [\href{http://arxiv.org/abs/1212.2636}{{\tt arXiv:1212.2636}}].

\bibitem{Gladyshev:2012xq}
A.~V. Gladyshev and D.~I. Kazakov, {\it {Is (Low Energy) SUSY Still Alive?}},
  in {\em {Proceedings, 2012 European School of High-Energy Physics (ESHEP
  2012): La Pommeraye, Anjou, France, June 06-19, 2012}}, pp.~107--159, 2014.
\newblock \href{http://arxiv.org/abs/1212.2548}{{\tt arXiv:1212.2548}}.

\bibitem{Kowalska:2012gs}
K.~Kowalska, S.~Munir, L.~Roszkowski, E.~M. Sessolo, S.~Trojanowski, and
  Y.-L.~S. Tsai, {\it {Constrained next-to-minimal supersymmetric standard
  model with a 126 GeV Higgs boson: A global analysis}},  {\em \prd} {\bf 87}
  (2013) 115010, [\href{http://arxiv.org/abs/1211.1693}{{\tt
  arXiv:1211.1693}}].

\bibitem{Mastercode12b}
O.~{Buchmueller}, R.~{Cavanaugh}, M.~{Citron}, A.~{De Roeck}, M.~J. {Dolan},
  J.~R. {Ellis}, H.~{Fl{\"a}cher}, S.~{Heinemeyer}, G.~{Isidori},
  J.~{Marrouche}, D.~{Mart{\'{\i}}nez Santos}, S.~{Nakach}, K.~A. {Olive},
  S.~{Rogerson}, F.~J. {Ronga}, K.~J. {de Vries}, and G.~{Weiglein}, {\it {The
  CMSSM and NUHM1 in light of 7 TeV LHC, $B_s\to\mu^+\mu-$ and XENON100 data}},
   {\em \epjc} {\bf 72} (Nov., 2012) 2243,
  [\href{http://arxiv.org/abs/1207.7315}{{\tt arXiv:1207.7315}}].

\bibitem{arXiv:1207.1839}
S.~Akula, P.~Nath, and G.~Peim, {\it {Implications of the Higgs Boson Discovery
  for mSUGRA}},  {\em \plb} {\bf 717} (2012) 188--192,
  [\href{http://arxiv.org/abs/1207.1839}{{\tt arXiv:1207.1839}}].

\bibitem{arXiv:1207.4846}
H.~Baer, V.~Barger, A.~Lessa, and X.~Tata, {\it {Discovery potential for SUSY
  at a high luminosity upgrade of LHC14}},  {\em \prd} {\bf 86} (2012) 117701,
  [\href{http://arxiv.org/abs/1207.4846}{{\tt arXiv:1207.4846}}].

\bibitem{Roszkowski12}
L.~{Roszkowski}, E.~M. {Sessolo}, and Y.-L.~S. {Tsai}, {\it {Bayesian
  implications of current LHC supersymmetry and dark matter detection searches
  for the constrained MSSM}},  {\em \prd} {\bf 86} (Nov., 2012) 095005,
  [\href{http://arxiv.org/abs/1202.1503}{{\tt arXiv:1202.1503}}].

\bibitem{SuperbayesHiggs}
C.~{Strege}, G.~{Bertone}, D.~G. {Cerde{\~n}o}, M.~{Fornasa}, R.~{Ruiz de
  Austri}, and R.~{Trotta}, {\it {Updated global fits of the cMSSM including
  the latest LHC SUSY and Higgs searches and XENON100 data}},  {\em \jcap} {\bf
  3} (Mar., 2012) 30, [\href{http://arxiv.org/abs/1112.4192}{{\tt
  arXiv:1112.4192}}].

\bibitem{Fittino12}
P.~{Bechtle}, T.~{Bringmann}, K.~{Desch}, H.~{Dreiner}, M.~{Hamer},
  C.~{Hensel}, M.~{Kr{\"a}mer}, N.~{Nguyen}, W.~{Porod}, X.~{Prudent},
  B.~{Sarrazin}, M.~{Uhlenbrock}, and P.~{Wienemann}, {\it {Constrained
  supersymmetry after two years of LHC data: a global view with Fittino}},
  {\em \jhep} {\bf 6} (June, 2012) 98,
  [\href{http://arxiv.org/abs/1204.4199}{{\tt arXiv:1204.4199}}].

\bibitem{Mastercode12}
O.~{Buchmueller}, R.~{Cavanaugh}, A.~{De Roeck}, M.~J. {Dolan}, J.~R. {Ellis},
  H.~{Fl{\"a}cher}, S.~{Heinemeyer}, G.~{Isidori}, D.~{Mart{\'{\i}}nez Santos},
  K.~A. {Olive}, S.~{Rogerson}, F.~J. {Ronga}, and G.~{Weiglein}, {\it
  {Supersymmetry in light of 1/fb of LHC data}},  {\em \epjc} {\bf 72} (Feb.,
  2012) 1878, [\href{http://arxiv.org/abs/1110.3568}{{\tt arXiv:1110.3568}}].

\bibitem{arXiv:1111.6098}
A.~Fowlie, A.~Kalinowski, M.~Kazana, L.~Roszkowski, and Y.~L.~S. Tsai, {\it
  {Bayesian Implications of Current LHC and XENON100 Search Limits for the
  Constrained MSSM}},  {\em \prd} {\bf 85} (2012) 075012,
  [\href{http://arxiv.org/abs/1111.6098}{{\tt arXiv:1111.6098}}].

\bibitem{Fittino}
P.~{Bechtle}, K.~{Desch}, M.~{Uhlenbrock}, and P.~{Wienemann}, {\it
  {Constraining SUSY models with Fittino using measurements before, with and
  beyond the LHC}},  {\em \epjc} {\bf 66} (Mar., 2010) 215--259,
  [\href{http://arxiv.org/abs/0907.2589}{{\tt arXiv:0907.2589}}].

\bibitem{Trotta08}
R.~{Trotta}, F.~{Feroz}, M.~{Hobson}, L.~{Roszkowski}, and R.~{Ruiz de Austri},
  {\it {The impact of priors and observables on parameter inferences in the
  constrained MSSM}},  {\em \jhep} {\bf 12} (2008) 24,
  [\href{http://arxiv.org/abs/0809.3792}{{\tt arXiv:0809.3792}}].

\bibitem{Fittino06}
P.~{Bechtle}, K.~{Desch}, and P.~{Wienemann}, {\it {Fittino, a program for
  determining MSSM parameters from collider observables using an iterative
  method}},  {\em \cpc} {\bf 174} (Jan., 2006) 47--70,
  [\href{http://arxiv.org/abs/hep-ph/0412012}{{\tt hep-ph/0412012}}].

\bibitem{arXiv:1608.02489}
K.~Kowalska, {\it {Phenomenological MSSM in light of new 13 TeV LHC data}},
  {\em Eur. Phys. J.} {\bf C76} (2016), no.~12 684,
  [\href{http://arxiv.org/abs/1608.02489}{{\tt arXiv:1608.02489}}].

\bibitem{arXiv:1507.07008}
G.~Bertone, F.~Calore, S.~Caron, R.~Ruiz, J.~S. Kim, R.~Trotta, and C.~Weniger,
  {\it {Global analysis of the pMSSM in light of the \textit{Fermi} GeV excess:
  prospects for the LHC Run-II and astroparticle experiments}},  {\em \jcap}
  {\bf 1604} (2016), no.~04 037, [\href{http://arxiv.org/abs/1507.07008}{{\tt
  arXiv:1507.07008}}].

\bibitem{Mastercode15}
E.~A. {Bagnaschi}, O.~{Buchmueller}, R.~{Cavanaugh}, M.~{Citron}, A.~{De
  Roeck}, M.~J. {Dolan}, J.~R. {Ellis}, H.~{Fl{\"a}cher}, S.~{Heinemeyer},
  G.~{Isidori}, S.~{Malik}, D.~{Mart{\'{\i}}nez Santos}, K.~A. {Olive},
  K.~{Sakurai}, K.~J. {de Vries}, and G.~{Weiglein}, {\it {Supersymmetric dark
  matter after LHC run 1}},  {\em \epjc} {\bf 75} (Oct., 2015) 500,
  [\href{http://arxiv.org/abs/1508.01173}{{\tt arXiv:1508.01173}}].

\bibitem{arXiv:1506.02499}
S.~S. AbdusSalam and L.~Velasco-Sevilla, {\it {Where to look for natural
  supersymmetry}},  {\em \prd} {\bf 94} (2016), no.~3 035026,
  [\href{http://arxiv.org/abs/1506.02499}{{\tt arXiv:1506.02499}}].

\bibitem{arXiv:1504.03260}
K.~J. de~Vries et~al., {\it {The pMSSM10 after LHC Run 1}},  {\em \epjc} {\bf
  75} (2015), no.~9 422, [\href{http://arxiv.org/abs/1504.03260}{{\tt
  arXiv:1504.03260}}].

\bibitem{Mastercode17}
E.~Bagnaschi et~al., {\it {Likelihood Analysis of the pMSSM11 in Light of LHC
  13-TeV Data}},  {\em Eur. Phys. J.} {\bf C78} (2018), no.~3 256,
  [\href{http://arxiv.org/abs/1710.11091}{{\tt arXiv:1710.11091}}].

\bibitem{Cheung:2012xb}
K.~Cheung, Y.-L.~S. Tsai, P.-Y. Tseng, T.-C. Yuan, and A.~Zee, {\it {Global
  Study of the Simplest Scalar Phantom Dark Matter Model}},  {\em \jcap} {\bf
  1210} (2012) 042, [\href{http://arxiv.org/abs/1207.4930}{{\tt
  arXiv:1207.4930}}].

\bibitem{Arhrib:2013ela}
A.~Arhrib, Y.-L.~S. Tsai, Q.~Yuan, and T.-C. Yuan, {\it {An Updated Analysis of
  Inert Higgs Doublet Model in light of the Recent Results from LUX, PLANCK,
  AMS-02 and LHC}},  {\em \jcap} {\bf 1406} (2014) 030,
  [\href{http://arxiv.org/abs/1310.0358}{{\tt arXiv:1310.0358}}].

\bibitem{Sming14}
S.~{Matsumoto}, S.~{Mukhopadhyay}, and Y.-L.~S. {Tsai}, {\it {Singlet Majorana
  fermion dark matter: a comprehensive analysis in effective field theory}},
  {\em \jhep} {\bf 10} (Oct., 2014) 155,
  [\href{http://arxiv.org/abs/1407.1859}{{\tt arXiv:1407.1859}}].

\bibitem{Chowdhury15}
D.~{Chowdhury} and O.~{Eberhardt}, {\it {Global fits of the two-loop
  renormalized Two-Higgs-Doublet model with soft Z $_{2}$ breaking}},  {\em
  \jhep} {\bf 11} (Nov., 2015) 52, [\href{http://arxiv.org/abs/1503.08216}{{\tt
  arXiv:1503.08216}}].

\bibitem{Liem16}
S.~{Liem}, G.~{Bertone}, F.~{Calore}, R.~R. {de Austri}, T.~M.~P. {Tait},
  R.~{Trotta}, and C.~{Weniger}, {\it {Effective field theory of dark matter: a
  global analysis}},  {\em \jhep} {\bf 9} (Sept., 2016) 77,
  [\href{http://arxiv.org/abs/1603.05994}{{\tt arXiv:1603.05994}}].

\bibitem{LikeDM}
X.~Huang, Y.-L.~S. Tsai, and Q.~Yuan, {\it {LikeDM: likelihood calculator of
  dark matter detection}},  {\em \cpc} {\bf 213} (2017) 252--263,
  [\href{http://arxiv.org/abs/1603.07119}{{\tt arXiv:1603.07119}}].

\bibitem{Banerjee:2016hsk}
S.~Banerjee, S.~Matsumoto, K.~Mukaida, and Y.-L.~S. Tsai, {\it {WIMP Dark
  Matter in a Well-Tempered Regime: A case study on Singlet-Doublets Fermionic
  WIMP}},  {\em \jhep} {\bf 11} (2016) 070,
  [\href{http://arxiv.org/abs/1603.07387}{{\tt arXiv:1603.07387}}].

\bibitem{Matsumoto:2016hbs}
S.~Matsumoto, S.~Mukhopadhyay, and Y.-L.~S. Tsai, {\it {Effective Theory of
  WIMP Dark Matter supplemented by Simplified Models: Singlet-like Majorana
  fermion case}},  {\em \prd} {\bf 94} (2016), no.~6 065034,
  [\href{http://arxiv.org/abs/1604.02230}{{\tt arXiv:1604.02230}}].

\bibitem{Cuoco:2016jqt}
A.~{Cuoco}, B.~{Eiteneuer}, J.~{Heisig}, and M.~{Kr{\"a}mer}, {\it {A global
  fit of the {$\gamma$}-ray galactic center excess within the scalar singlet
  Higgs portal model}},  {\em \jcap} {\bf 6} (June, 2016) 050,
  [\href{http://arxiv.org/abs/1603.08228}{{\tt arXiv:1603.08228}}].

\bibitem{Cacchio:2016qyh}
V.~Cacchio, D.~Chowdhury, O.~Eberhardt, and C.~W. Murphy, {\it {Next-to-leading
  order unitarity fits in Two-Higgs-Doublet models with soft $\mathbb{Z}_2$
  breaking}},  {\em \jhep} {\bf 11} (2016) 026,
  [\href{http://arxiv.org/abs/1609.01290}{{\tt arXiv:1609.01290}}].

\bibitem{BertoneUED}
G.~{Bertone}, K.~{Kong}, R.~R. {de Austri}, and R.~{Trotta}, {\it {Global fits
  of the minimal universal extra dimensions scenario}},  {\em \prd} {\bf 83}
  (Feb., 2011) 036008, [\href{http://arxiv.org/abs/1010.2023}{{\tt
  arXiv:1010.2023}}].

\bibitem{Chiang:2018cgb}
C.-W. Chiang, G.~Cottin, and O.~Eberhardt, {\it {Global fits in the
  Georgi-Machacek model}},  {\em \prd} {\bf 99} (2019), no.~1 015001,
  [\href{http://arxiv.org/abs/1807.10660}{{\tt arXiv:1807.10660}}].

\bibitem{hepfit}
J.~De~Blas et~al., {\it {$\texttt{HEPfit}$: a Code for the Combination of
  Indirect and Direct Constraints on High Energy Physics Models}},  {\em \epjc}
  {\bf 80} (2019) 456, [\href{http://arxiv.org/abs/1910.14012}{{\tt
  arXiv:1910.14012}}].

\bibitem{Matsumoto:2018acr}
S.~Matsumoto, Y.-L.~S. Tsai, and P.-Y. Tseng, {\it {Light Fermionic WIMP Dark
  Matter with Light Scalar Mediator}},  {\em \jhep} {\bf 07} (2019) 050,
  [\href{http://arxiv.org/abs/1811.03292}{{\tt arXiv:1811.03292}}].

\bibitem{1306.2144}
F.~{Feroz}, M.~P. {Hobson}, E.~{Cameron}, and A.~N. {Pettitt}, {\it {Importance
  Nested Sampling and the MultiNest Algorithm}},
  \href{http://arxiv.org/abs/1306.2144}{{\tt arXiv:1306.2144}}.

\bibitem{0809.3437}
F.~{Feroz}, M.~P. {Hobson}, and M.~{Bridges}, {\it {MULTINEST: an efficient and
  robust Bayesian inference tool for cosmology and particle physics}},  {\em
  \mnras} {\bf 398} (Oct., 2009) 1601--1614,
  [\href{http://arxiv.org/abs/0809.3437}{{\tt arXiv:0809.3437}}].

\bibitem{0704.3704}
F.~{Feroz} and M.~P. {Hobson}, {\it {Multimodal nested sampling: an efficient
  and robust alternative to Markov Chain Monte Carlo methods for astronomical
  data analyses}},  {\em \mnras} {\bf 384} (Feb., 2008) 449--463,
  [\href{http://arxiv.org/abs/0704.3704}{{\tt arXiv:0704.3704}}].

\bibitem{Dunkley05}
J.~{Dunkley}, M.~{Bucher}, P.~G. {Ferreira}, K.~{Moodley}, and C.~{Skordis},
  {\it {Fast and reliable Markov chain Monte Carlo technique for cosmological
  parameter estimation}},  {\em \mnras} {\bf 356} (Jan., 2005) 925--936,
  [\href{http://arxiv.org/abs/astro-ph/0405462}{{\tt astro-ph/0405462}}].

\bibitem{GreAT}
A.~{Putze} and L.~{Derome}, {\it {The Grenoble Analysis Toolkit (GreAT)-A
  statistical analysis framework}},  {\em Physics of the Dark Universe} {\bf 5}
  (Dec., 2014) 29--34.

\bibitem{Hastings}
W.~K. Hastings, {\it Monte carlo sampling methods using markov chains and their
  applications},  {\em Biometrika} {\bf 57} (1970), no.~1 97--109.

\bibitem{vanBeekveld:2019tqp}
M.~van Beekveld, S.~Caron, and R.~Ruiz~de Austri, {\it {The current status of
  fine-tuning in supersymmetry}},  {\em JHEP} {\bf 01} (2020) 147,
  [\href{http://arxiv.org/abs/1906.10706}{{\tt arXiv:1906.10706}}].

\bibitem{Athron:2017yua}
{\bf \GB} Collaboration, P.~Athron et~al., {\it {A global fit of the MSSM with
  GAMBIT}},  {\em Eur. Phys. J. C} {\bf 77} (2017), no.~12 879,
  [\href{http://arxiv.org/abs/1705.07917}{{\tt arXiv:1705.07917}}].

\bibitem{Hogg:2017akh}
D.~W. Hogg and D.~Foreman-Mackey, {\it {Data analysis recipes: Using Markov
  Chain Monte Carlo}},  {\em Astrophys. J. Suppl.} {\bf 236} (2018), no.~1 11,
  [\href{http://arxiv.org/abs/1710.06068}{{\tt arXiv:1710.06068}}].

\bibitem{Skilling:2006gxv}
J.~Skilling, {\it {Nested sampling for general Bayesian computation}},  {\em
  Bayesian Analysis} {\bf 1} (2006), no.~4 833--859.

\bibitem{Kvellestad:2019vxm}
A.~Kvellestad, P.~Scott, and M.~White, {\it {GAMBIT and its Application in the
  Search for Physics Beyond the Standard Model}},
  \href{http://arxiv.org/abs/1912.04079}{{\tt arXiv:1912.04079}}.

\bibitem{Hoof:2018ieb}
S.~Hoof, F.~Kahlhoefer, P.~Scott, C.~Weniger, and M.~White, {\it {Axion global
  fits with Peccei-Quinn symmetry breaking before inflation using GAMBIT}},
  {\em JHEP} {\bf 03} (2019) 191, [\href{http://arxiv.org/abs/1810.07192}{{\tt
  arXiv:1810.07192}}]. [Erratum: JHEP 11, 099 (2019)].

\bibitem{Athron:2018hpc}
{\bf \GB} Collaboration, P.~Athron et~al., {\it {Global analyses of Higgs
  portal singlet dark matter models using GAMBIT}},  {\em Eur. Phys. J. C} {\bf
  79} (2019), no.~1 38, [\href{http://arxiv.org/abs/1808.10465}{{\tt
  arXiv:1808.10465}}].

\bibitem{Athron:2017qdc}
{\bf \GB} Collaboration, P.~Athron et~al., {\it {Global fits of GUT-scale SUSY
  models with GAMBIT}},  {\em Eur. Phys. J. C} {\bf 77} (2017), no.~12 824,
  [\href{http://arxiv.org/abs/1705.07935}{{\tt arXiv:1705.07935}}].

\bibitem{Athron:2018vxy}
{\bf \GB} Collaboration, P.~Athron et~al., {\it {Combined collider constraints
  on neutralinos and charginos}},  {\em Eur. Phys. J. C} {\bf 79} (2019), no.~5
  395, [\href{http://arxiv.org/abs/1809.02097}{{\tt arXiv:1809.02097}}].

\bibitem{ScannerBit}
{\bf \GB Scanner Workgroup} Collaboration, G.~D. {Martinez}, J.~{McKay},
  B.~{Farmer}, P.~{Scott}, E.~{Roebber}, A.~{Putze}, and J.~{Conrad}, {\it
  {Comparison of statistical sampling methods with ScannerBit, the GAMBIT
  scanning module}},  {\em \epjc} {\bf 77} (May, 2017) 761,
  [\href{http://arxiv.org/abs/1705.07959}{{\tt arXiv:1705.07959}}].

\bibitem{StornPrice95}
R.~Storn and K.~Price, {\it Differential evolution: A simple and efficient
  heuristic for global optimization over continuous spaces},  {\em Journal of
  Global Optimization} {\bf 11} (1997), no.~4 341--359.

\bibitem{Price05wholebook}
K.~Price, R.~M. Storn, and J.~A. Lampinen, {\em Differential evolution: a
  practical approach to global optimization}.
\newblock Springer, 2005.

\bibitem{DasSuganthan11}
S.~Das and P.~Suganthan, {\it Differential evolution: A survey of the
  state-of-the-art},  {\em Evolutionary Computation, IEEE Transactions on} {\bf
  15} (Feb, 2011) 4--31.

\bibitem{Price13}
K.~Price, {\it Differential evolution},  in {\em Handbook of Optimization}
  (I.~Zelinka, V.~Sn\'{a}\v{s}el, and A.~Abraham, eds.), vol.~38 of {\em
  Intelligent Systems Reference Library}, pp.~187--214.
\newblock Springer Berlin Heidelberg, 2013.

\bibitem{Brest06}
J.~Brest, S.~Greiner, B.~Boskovic, M.~Mernik, and V.~Zumer, {\it Self-adapting
  control parameters in differential evolution: A comparative study on
  numerical benchmark problems},  {\em Evolutionary Computation, IEEE
  Transactions on} {\bf 10} (Dec, 2006) 646--657.

\bibitem{Athron:2017ard}
{\bf \GB} Collaboration, P.~Athron et~al., {\it {GAMBIT: The Global and Modular
  Beyond-the-Standard-Model Inference Tool}},  {\em Eur. Phys. J. C} {\bf 77}
  (2017), no.~11 784, [\href{http://arxiv.org/abs/1705.07908}{{\tt
  arXiv:1705.07908}}]. [Addendum: Eur.Phys.J.C 78, 98 (2018)].

\bibitem{jde}
J.~Brest, S.~Greiner, B.~Bošković, M.~Mernik, and V.~Zumer, {\it
  Self-adapting control parameters in differential evolution: a comparative
  study on numerical benchmark problems},  {\em Evolutionary Computation}
  (2006).

\bibitem{ide}
S.~M. Elsayed, R.~A. Sarker, and D.~L. Essam {\em Evolutionary Computation}
  (2011).

\bibitem{pygmo}
D.~I. F.~Biscani, {\it Pygmo (1.1.7)},  2015.

\bibitem{488968}
J.~{Kennedy} and R.~{Eberhart}, {\it Particle swarm optimization},  in {\em
  Proceedings of ICNN'95 - International Conference on Neural Networks},
  vol.~4, pp.~1942--1948 vol.4, Nov, 1995.

\bibitem{Bonyadi}
M.~R. Bonyadi and Z.~Michalewicz, {\it Particle swarm optimization for single
  objective continuous space problems: A review},  {\em Evolutionary
  Computation} {\bf 25} (2017), no.~1 1--54.

\bibitem{Hansen:2006CMAES}
N.~Hansen, {\em The CMA Evolution Strategy: A Comparing Review}, pp.~75--102.
\newblock Springer Berlin Heidelberg, Berlin, Heidelberg, 2006.

\bibitem{SnoekPBO}
J.~Snoek, H.~Larochelle, and R.~Adams, {\it {Practical Bayesian optimization of
  machine learning algorithms}},  \href{http://arxiv.org/abs/1206.2944v}{{\tt
  arXiv:1206.2944v}}.

\bibitem{BullCGO}
A.~Bull, {\it {Convergence rates of efficient global optimization algorithms}},
   \href{http://arxiv.org/abs/1101.3501v}{{\tt arXiv:1101.3501v}}.

\bibitem{RasmGPML}
C.~Rasmussen and C.~Williams, {\em Gaussian Processes for Machine Learning}.
\newblock Massachusetts Institute of Technology, 2006.

\bibitem{eriksson2019scalable}
D.~Eriksson, M.~Pearce, J.~R. Gardner, R.~Turner, and M.~Poloczek, {\it
  Scalable global optimization via local bayesian optimization},  {\em arXiv
  preprint arXiv:1910.01739} (2019).

\bibitem{gworef}
S.~Mirjalili, S.~Mirjalili, and A.~Lewis, {\it Grey wolf optimizer},  {\em
  Advances in Engineering Software} {\bf 69} (1, 2014) 46--61.

\bibitem{abc}
D.~Karaboga, {\it An idea based on honey bee swarm for numerical optimization},
   2005.

\bibitem{AKAY2012120}
B.~Akay and D.~Karaboga, {\it A modified artificial bee colony algorithm for
  real-parameter optimization},  {\em Information Sciences} {\bf 192} (2012)
  120 -- 142. Swarm Intelligence and Its Applications.

\bibitem{MERNIK2015115}
M.~Mernik, S.-H. Liu, D.~Karaboga, and M.~Črepinšek, {\it On clarifying
  misconceptions when comparing variants of the artificial bee colony algorithm
  by offering a new implementation},  {\em Information Sciences} {\bf 291}
  (2015) 115 -- 127.

\bibitem{1232326}
J.~H. Kotecha and P.~M. Djuric, {\it Gaussian particle filtering},  {\em IEEE
  Transactions on Signal Processing} {\bf 51} (Oct, 2003) 2592--2601.

\bibitem{AMPGO}
L.~Lasdon, A.~Duarte, F.~Glover, M.~Laguna, and R.~Marti, {\it Adaptive memory
  programming for constrained global optimization},  {\em Computers \&
  Operations Research} {\bf 37} (08, 2010) 1500--1509.

\bibitem{AMPGO_TabuTunneling}
A.~Levy and S.~Gómez, {\em The tunneling method applied to global
  optimization}, pp.~213--244.
\newblock 01, 1985.

\bibitem{Byrd94alimited-memory}
R.~H. Byrd, P.~Lu, J.~Nocedal, and C.~Zhu, {\it A limited-memory algorithm for
  bound constrained optimization},  {\em SIAM JOURNAL ON SCIENTIFIC COMPUTING}
  {\bf 16} (1994) 1190--1208.

\bibitem{AMPGO_Site}
``Ampgo analysis.'' \url{http://infinity77.net/global_optimization/ampgo.html}.
\newblock Accessed: 2020-01-04.

\bibitem{Athron_2017}
P.~Athron, C.~Balázs, T.~Bringmann, A.~Buckley, M.~Chrząszcz, J.~Conrad,
  J.~M. Cornell, L.~A. Dal, J.~Edsjö, and et~al., {\it A global fit of the
  mssm with gambit},  {\em The European Physical Journal C} {\bf 77} (Dec,
  2017).

\bibitem{Brooijmans:2020yij}
G.~Brooijmans et~al., {\it {Les Houches 2019 Physics at TeV Colliders: New
  Physics Working Group Report}},  in {\em {11th Les Houches Workshop on
  Physics at TeV Colliders}: {PhysTeV Les Houches}}, 2, 2020.
\newblock \href{http://arxiv.org/abs/2002.12220}{{\tt arXiv:2002.12220}}.

\bibitem{Buckley:2011kc}
A.~Buckley, A.~Shilton, and M.~White, {\it {Fast supersymmetry phenomenology at
  the Large Hadron Collider using machine learning techniques}},  {\em Comput.
  Phys. Commun.} {\bf 183} (2012) 960--970,
  [\href{http://arxiv.org/abs/1106.4613}{{\tt arXiv:1106.4613}}].

\bibitem{klambauer2017selfnormalizing}
G.~Klambauer, T.~Unterthiner, A.~Mayr, and S.~Hochreiter, {\it Self-normalizing
  neural networks},  2017.

\bibitem{kingma2017adam}
D.~P. Kingma and J.~Ba, {\it Adam: A method for stochastic optimization},
  2017.

\end{thebibliography}\endgroup

\end{document}